\documentclass{emulateapj}
\usepackage{lscape}

\def\kms{\ifmmode{\rm km\thinspace s^{-1}}\else km\thinspace s$^{-1}$\fi}
\def\ms{\ifmmode{\rm m\thinspace s^{-1}}\else m\thinspace s$^{-1}$\fi}
\def\hip{{\it Hipparcos\/}}
\def\today{\number\year\space \ifcase\month\or  January\or February\or
        March\or April\or May\or June\or July\or August\or
September\or
        October\or November\or December\fi\space \number\day}

\shortauthors{Torres et al.}
\shorttitle{Improved parameters}

\begin{document}

\journalinfo{Accepted for publication in The Astrophysical Journal}

%\email{********* Draft Version \today\ *********}

\title{Improved parameters for extrasolar transiting planets}

\author{
Guillermo Torres\altaffilmark{1},
Joshua N.\ Winn\altaffilmark{2}, and
Matthew J.\ Holman\altaffilmark{1}
}

\altaffiltext{1}{Harvard-Smithsonian Center for Astrophysics, 60
Garden St., Cambridge, MA 02138, USA}

\altaffiltext{2}{Department of Physics, and Kavli Institute for
Astrophysics and Space Research, Massachusetts Institute of
Technology, Cambridge, MA 02139, USA}

\email{gtorres@cfa.harvard.edu}

\begin{abstract}

We present refined values for the physical parameters of transiting
exoplanets, based on a self-consistent and uniform analysis of transit
light curves and the observable properties of the host
stars. Previously it has been difficult to interpret the ensemble
properties of transiting exoplanets, because of the widely different
methodologies that have been applied in individual cases.
Furthermore, previous studies often ignored an important constraint on
the mean stellar density that can be derived directly from the light
curve. The main contributions of this work are \emph{i}) a critical
compilation and error assessment of all reported values for the
effective temperature and metallicity of the host stars; \emph{ii})
the application of a consistent methodology and treatment of errors in
modeling the transit light curves; and \emph{iii}) more accurate
estimates of the stellar mass and radius based on stellar evolution
models, incorporating the photometric constraint on the stellar
density.  We use our results to revisit some previously proposed
patterns and correlations within the ensemble.  We confirm the
mass-period correlation, and we find evidence for a new pattern within
the scatter about this correlation: planets around metal-poor stars
are more massive than those around metal-rich stars at a given orbital
period.  Likewise, we confirm the proposed dichotomy of planets
according to their Safronov number, and we find evidence that the
systems with small Safronov numbers are more metal-rich on
average. Finally, we confirm the trend that led to the suggestion that
higher-metallicity stars harbor planets with a greater heavy-element
content.

\end{abstract}


\section{Introduction}
\label{sec:introduction}

The transiting exoplanets are only a small subset of all the known
planets orbiting other stars, but they hold tremendous promise for
deepening our understanding of planetary formation, structure, and
evolution. Observations of transits and occultations\footnote{The word
\emph{transit} is sometimes assumed in the field of exoplanet research
to be synonymous with \emph{eclipse}. In reality, it has a more
restricted meaning and has long been used in the eclipsing binary
field to describe an eclipse of the larger object by the smaller one.
The term \emph{occultation} is used to refer to the passage of the
smaller object (in this case, the planet) behind the larger one (the
star) \citep[see, e.g.,][]{Popper:76}. To avoid confusion, we advocate
that \emph{occultation} or \emph{secondary eclipse} are preferable to
neologisms such as ``secondary transit'' or ``anti-transit''.}  (along
with the spectroscopic orbit of the host star) not only allow one to
measure the mass and radius of the planet, but also provide
opportunities to measure the stellar spin-orbit alignment
\citep{Queloz:00, Winn:07a}, the planetary brightness temperature
\citep{Charbonneau:05, Deming:05}, the planetary day-night temperature
difference \citep{Knutson:07}, and even absorption lines of planetary
atmospheric constituents \citep{Charbonneau:02, Vidal-Madjar:04,
Tinetti:07}. These and other observations have been accompanied by
theoretical progress in modeling the physical processes in the
planetary interiors and atmospheres, as well as the planets'
interactions with their parent stars. This rapid progress has been
stimulated in no small measure by the remarkable diversity of planet
characteristics that has been found among the members of the
transiting ensemble.

The accuracy and precision with which the properties of the planet can
be derived from transit data depend strongly on whatever measurements
and assumptions are made regarding the host star. For example, for a
given value of the transit depth, the inferred planetary radius scales
in proportion to the assumed stellar radius; and for a given
spectroscopic orbit of the host star, the inferred planetary mass
scales as the two-thirds power of the stellar mass.  In the literature
on transiting planets, a wide variety of methods have been used to
estimate the radius and mass of the parent star, ranging from simply
looking them up in a table of average stellar properties as a function
of spectral type, all the way to fitting detailed stellar evolutionary
models constrained by the luminosity, effective temperature, and other
observations that may be available for the star. As a result, the
ensemble of planet properties at our disposal is inhomogeneous, and in
many cases the uncertainty of those determinations is dominated by
systematic errors in the stellar parameters that are treated
differently by different investigators.

This situation is unfortunate because it hinders our ability to gauge
the reliability of any patterns that are discerned among the ensemble
properties of transiting exoplanets. With 23 systems that have been
reported in the literature, this subfield should be poised for the
transition from a handful of results to a large and diverse enough
sample for meaningful general conclusions to be drawn, but the
heterogeneity of reported results is clearly an obstacle.

It may be surprising that we are limited in many cases by our
knowledge of the properties of the parent stars; one would think that
stellar physics is a ``solved problem'' in comparison to exoplanetary
physics. However it must be remembered that the host stars are usually
isolated (and therefore no dynamical mass measurement is possible),
and that many of the host stars are distant enough that they do not
even have measured parallaxes.  Of the 23 cases in the literature,
five have \hip\ parallaxes \citep{Perryman:97}.  For those few it is
fairly straightforward to estimate the stellar properties, but for the
other systems, more indirect methods have been used.  These indirect
methods often rely on the value of the stellar surface gravity that is
derived by measuring the depths and shapes of gravity-sensitive
absorption lines in the stellar spectrum.  This is a notoriously
difficult measurement and the result is often strongly correlated with
other parameters that affect the spectrum.  Recently, however,
\cite{Sozzetti:07}, Holman et al.~(2007), and others demonstrated that
it is possible to do better by using the information about the mean
stellar density that is encoded in the transit light curve. This
information was typically overlooked prior to these studies.

The study presented in this paper was motivated by the desire for a
homogeneous analysis, and by the desire to take advantage of the
photometric estimates of the stellar mean density.  We have revisited
the determination of the stellar parameters for all of the transiting
planets that have been reported in the literature.  We have taken the
opportunity to merge all existing measurements of the atmospheric
parameters (mainly the effective temperature and metallicity) with the
goal of presenting the best possible values.  We have chosen a uniform
method for analyzing photometric data and have re-analyzed existing
light curves where necessary to provide homogeneity.  Our hope was
that by applying these procedures across the board, we and other
investigators could view the ensemble properties with greater clarity
and uncover any interesting clues the transiting planets might provide
us about the origin, structure, and evolution of exoplanets.

This paper is organized as follows. \S\,\ref{sec:stellar} describes
the procedures by which we estimated the stellar properties, using the
available spectroscopic and photometric datasets, and a particular set
of theoretical stellar evolution models. As a check on the models,
\S\,\ref{sec:modchecks} compares the results of a subset of our
calculations with those derived from a different set of evolutionary
models. \S\,\ref{sec:obschecks} investigates alternate ways of
estimating the stellar properties. \S\,\ref{sec:gj436} deals with
GJ~436, which needs special treatment because the host star has such a
lower mass than the other host stars. \S\,\ref{sec:results} presents
the final results for the planetary parameters.
\S\,\ref{sec:discussion} uses the new results to check on some of the
previously proposed correlations among the properties of the
transiting ensemble, and \S\,\ref{sec:remarks} provides final remarks.
The Appendix lists the data sets and other issues that are particular
to each system.

\section{Determining the stellar properties}
\label{sec:stellar}

For a transiting planet, the basic data are the spectroscopic orbit of
the star (radial-velocity curve), and the photometric observations of
transits (light curve).  With such data, the planetary mass and radius
cannot be determined independently of the stellar properties.  The
radial-velocity curve can be used to determine
\begin{equation}
M_p \sin i = 4.919 \times 10^{-3} P^{1/3} (1-e^2)^{1/2} K_{\star}
\left[(M_{\star} + M_p)/M_{\odot}\right]^{2/3}
\end{equation}
(in units of the mass of Jupiter), where $i$ is the inclination angle
of the orbit, $P$ is the orbital period in days, $e$ is the
eccentricity, and $K_{\star}$ the velocity semi-amplitude of the star
in \ms.  Even when $\sin i$ is known precisely, the value of the
planetary mass $M_p$ that is derived from the data will scale as
$M_\star^{2/3}$, where $M_{\star}$ is the mass of the star.
Meanwhile, the light curve does not immediately yield $R_p$, the
planetary radius; rather, the ratio of the radii $R_p/R_{\star}$ is
pinned down through the depth of the transit.

The stellar mass and radius are usually inferred indirectly, from an
analysis of high-resolution spectra of the star with the aid of model
atmospheres, followed by a comparison of the atmospheric parameters
with stellar evolution models. The latter step is performed somewhat
differently by different authors depending on the observational
constraints available. For this work we have compiled and critically
reviewed all of the available information regarding the atmospheric
properties of the host stars. This effort is described in
\S\,\ref{sec:atmospheric}.

The approach adopted in this paper makes use of information from the
light curves of transiting planets to constrain the mean density of
the star, $\rho_{\star}$, following \cite{Sozzetti:07}. As described
therein, the quantity $a/R_{\star}$ (the planet-star separation $a$ in
units of the stellar radius) is directly related to the mean stellar
density (hereafter, simply the density), and can be derived from the
photometry without much knowledge about the star (see below). The only
dependence $a/R_{\star}$ has on the stellar properties is through the
limb-darkening coefficients, which is typically a second-order effect
(see \S\,\ref{sec:lightcurves}).

Many of the published light curve analyses do not report the value of
$a/R_{\star}$ explicitly, and it is difficult or impossible to
reconstruct its value accurately from the published information. It is
especially difficult to obtain a good measure of the true uncertainty
in $a/R_\star$ from the published information.  For this reason, and
for the sake of homogeneity, we have re-analyzed many of the
high-quality photometric time-series available to us for all
transiting planets. We describe this effort in
\S\,\ref{sec:lightcurves}.

\subsection{Atmospheric parameters}
\label{sec:atmospheric}

High-resolution spectroscopic studies have been undertaken for almost
all of the parent stars of the known transiting planets.  The basic
products of these studies are measurements of the effective
temperature $T_{\rm eff}$, iron abundance [Fe/H], and surface gravity
$\log g$. The quoted precision of these determinations varies widely,
depending on the signal-to-noise ratio and resolution of the
observations, the modeling techniques employed, and the attitude taken
toward systematic errors.

For this work we have relied on published determinations, rather than
any new spectroscopic data.  In many cases, a given transiting system
has been described by more than one analysis, and they do not
necessarily agree. In those cases, rather than arbitrarily choosing
the most recent study, or the one with the smallest uncertainties, we
critically examined all of the available studies and combined them.
Our choices are documented in the Appendix. We were guided by our own
experience and increased the error estimates whenever they seemed
optimistic. For example, many of the automated spectroscopic analysis
tools in common use today return formal uncertainties for the
effective temperatures that are only a few tens of degrees Kelvin; for
one of the transiting systems the quoted error was only 13~K. There is
ample literature on the subject of the absolute effective temperature
scale and the systematic errors inherent in placing any single system
on such a scale. A recent investigation by \cite{Ramirez:04} has shown
that there are still differences of order 100~K between temperatures
derived from the spectroscopic condition of excitation equilibrium and
from the Infrared Flux Method \citep[IRFM,][]{Blackwell:77,
Blackwell:80}. Other studies have indicated discrepancies of 50--100~K
between different temperature scales \citep{Ramirez:05,
Casagrande:06}, and discussed at length possible sources for these
errors.  In light of these findings, we considered it prudent to adopt
temperature errors no smaller than 50~K for our study, and then only
when there are several independent and consistent determinations or
other evidence supporting that level of accuracy. Similar concerns
hold for the spectroscopic surface gravities, although we do not
actually make use of them in this work, as described below. For the
spectroscopic metallicities, we adopted a minimum uncertainty of 0.05
dex (for the best cases with multiple independent measurements), even
though smaller errors have occasionally been reported for individual
analyses. This is mainly because of the strong correlations present
among [Fe/H], $T_{\rm eff}$, and $\log g$ \citep[e.g.,][]{Buzzoni:01},
as well as some evidence for systematic differences between different
groups \citep[see, e.g.,][]{Santos:04, Fischer:05, Gonzalez:07}. In
the Appendix we provide a complete listing of the sources we have
drawn from in each case. The values of $T_{\rm eff}$, [Fe/H], and
$\log g$ finally adopted for all systems are listed in
Table~\ref{tab:atmospheric}.  Some degree of non-uniformity in these
quantities is unavoidable due to the variety of procedures used by
different authors, but we believe they represent the best available
set for the parent stars based on current knowledge.

\subsection{Light curve fits}
\label{sec:lightcurves}

For each system, we examined the highest-quality transit photometry
available to us in order to determine the key parameters
$R_p/R_\star$, $a/R_\star$, and $i$. Our methodology is described
below, and the details of the data sets that were used in each case
are given in the Appendix. In some cases, the published determinations
of $R_p/R_\star$, $a/R_\star$, and $i$ matched our own methodology
very closely, and we simply adopted the values from the literature;
these cases are also specified in the Appendix.

In the absence of limb darkening, the four primary observables in a
transit light curve are the midtransit time, the depth, the total
duration, and the partial-phase duration (ingress or egress). A
sequence of measured midtransit times usually leads to a very precise
determination of the orbital period $P$. The depth is equal to the
planet-to-star radius ratio, $(R_p/R_\star)^2$. At fixed $P$, the
parameters $i$ and $a/R_\star$ can be written in terms of the total
and partial durations through an application of Kepler's Law
\citep[see, e.g.,][]{Seager:03}. With limb darkening, however, there
is no longer a well-defined depth or partial duration; an accurate
light curve model must include a realistic intensity distribution
across the stellar disk, which will depend on the stellar temperature,
surface gravity, and metallicity. Thus, to some degree, the
determination of $R_p/R_\star$, $a/R_\star$, and $i$ must be
accompanied by some assumptions about the stellar properties.

Our procedure was as follows. We modeled each system using a two-body
Keplerian orbit. The star has mass $M_\star$ and radius $R_\star$, and
the planet has mass $M_p$ and radius $R_p$. The orbit has period $P$,
eccentricity $e$, argument of pericenter $\omega$, and inclination
$i$. For our purpose the uncertainty in $P$ was completely negligible;
we fixed $P$ at the most precisely determined value in the literature.
In almost all cases, the radial-velocity data are consistent with a
circular orbit, and we assumed $e=0$ exactly. For HAT-P-2 (HD~147506)
we fixed the values of $e$ and $\omega$ at those that have been
derived from radial-velocity data (and in the end we verified that
changing these parameters by 1$\sigma$ does not significantly affect
any of our final results). The initial condition is specified by a
particular midtransit time $T_c$. When the sky projections of the star
and planet do not overlap, the model flux is unity. When they do
overlap we use the analytic formulas of \cite{Mandel:02} to compute
the integral of the intensity over the unobscured portion of the
stellar disk, assuming a quadratic limb-darkening law.

To arrive at a self-consistent solution including limb darkening, we
began with initial values for $M_\star$ and $M_p$ from the literature.
We also chose values for the stellar $T_{\rm eff}$, $\log g$, and
metallicity, as described in the previous section. Then we adopted
limb-darkening coefficients based on those stellar parameters, by
interpolating the tables of \cite{Claret:00, Claret:04} for the
appropriate bandpass. At this point we estimated all of the remaining
parameters (using the procedure described in the next paragraph) and
used the result for $a/R_\star$ to update the determination of the
stellar properties. Then, a new photometric parameter estimation was
performed, with revised values of $M_\star$, $M_p$ and the
limb-darkening coefficients, and so forth. This procedure converged
after two or three iterations, in the sense that further iterations
changed none of the parameters by more than about a tenth of the
statistical error.

The parameter estimations were carried out with a Markov Chain Monte
Carlo algorithm (MCMC; see, e.g., \cite{Tegmark:04} for applications
to cosmological data, \cite{Ford:05} for radial-velocity data, and
\cite{Holman:06} or \cite{Burke:07} for a similar approach to transit
fitting). It is based on the goodness-of-fit statistic
\begin{equation}
\chi^2 =
\sum_{j=1}^{N_f}
\left[
\frac{f_j({\mathrm{obs}}) - f_j({\mathrm{calc}})}{\sigma_j}
\right]^2
,
\label{eq:chi2}
\end{equation}
where $f_j$(obs) is the flux observed at time $j$ and $\sigma_j$
controls the weights of the data points, and $f_j$(calc) is the flux
calculated with our model. In the MCMC algorithm, a stochastic process
is used to create a sequence of points in parameter space whose
density approximates the joint {\it a posteriori} probability density
for all parameters. One begins with an initial point and iterates a
jump function, which in our case was the addition of a Gaussian random
number (``perturbation'') to a randomly chosen parameter. If the new
point has a lower $\chi^2$, the jump is executed; if not, the jump is
executed with probability $\exp(-\Delta\chi^2/2)$.  We adjusted the
sizes of the perturbations until approximately $\sim$25\% of jumps are
executed for each parameter.

For the data weights $\sigma_j$, we used the observed standard
deviation of the out-of-transit data ($\sigma_1$), multiplied by a
factor $\beta\geq 1$ to account at least approximately for
time-correlated errors (``red'' noise) which are often significant for
ground-based data. We chose $\beta$ as follows. The key timescale is
the partial-phase duration, because the limiting error in the
determination of $a/R_\star$ and $i$ is generally the fractional error
in the partial-phase duration.  We averaged the out-of-transit data
over this timescale, with each time bin consisting of $N$ points
depending on the cadence of observations, and then calculated the
standard deviation of the binned data, $\sigma_N$. Finally we set
$\beta = \sigma_N * \sqrt{N} / \sigma_1$. With white noise only, we
would observe $\beta=1$, but in practice $\beta>1$ because the number
of effectively independent data points is smaller than the actual
number of data points. For this paper we deliberately chose to analyze
only those data for which $\beta < 2$.

For each parameter, we took the mode of the MCMC distribution after
marginalizing over all other parameters to be the ``best value.'' We
defined the 68\% confidence limits $p_{\rm lo}$ and $p_{\rm hi}$ as
the values for which the integral of the distribution between $p_{\rm
lo}$ and $p_{\rm hi}$ is 0.68, and the integrals from the minimum
value to $p_{\rm lo}$ and from $p_{\rm hi}$ to the maximum value were
both $0.16 = (1.00 - 0.68)/2$. In some cases the mode was all the way
at one end of the probability distribution; in particular there were
several cases in which $i=90\arcdeg$ was the mode of the inclination
distribution. In those cases we report $p_{\rm hi}$ as $90\arcdeg$ and
$p_{\rm lo}$ as the value for which the integral from zero to $p_{\rm
lo}$ was $(1.00-0.68) = 0.32$. The final results are given in
Table~\ref{tab:lcfits}, including the stellar density $\rho_{\star}$
computed directly from $a/R_{\star}$ and the period.

\subsection{Stellar masses, radii, luminosities, surface gravities, and
ages}
\label{sec:masses}

The fundamental parameters for the host stars are derived here using
stellar evolution models. We rely on the spectroscopically determined
$T_{\rm eff}$ and [Fe/H], and we require also an indicator of
luminosity ($L_{\star}$) or some other measure of evolution. Since
most of these stars lack a parallax measurement, the spectroscopic
surface gravity has often been used in the past as a luminosity
indicator.  The effect of $\log g$ on the spectral lines is relatively
subtle, and strong correlations between $\log g$ and both temperature
and metallicity make these determinations challenging.
\cite{Seager:03} have pointed out that an important property intrinsic
to the star, the \emph{density}, is encoded in the transit light
curves mainly through its dependence on the transit duration.  As
shown there, and more explicitly by \cite{Sozzetti:07}, the density is
directly related to $a/R_{\star}$, one of the parameters often solved
for in modeling the photometry.  This quantity can typically be
determined more precisely than $\log g$ and it is highly sensitive to
the degree of evolution of the star, i.e., to its size. Thus, it
serves as a better proxy for luminosity in many cases.  We illustrate
this below.

To determine the stellar mass and radius, and other relevant
properties of the host stars, we follow the procedure described by
\cite{Sozzetti:07} with minor improvements, and compare model
isochrones directly with the measured values of $T_{\rm eff}$, [Fe/H],
and $a/R_{\star}$. The latter can be calculated from the models as
\begin{equation}
\label{eq:aR}
{a\over R_\star} = \left({G\over 4\pi^2}\right)^{1/3} {P^{2/3}\over
R_\star} \left(M_\star + M_p\right)^{1/3}~,
\end{equation}
where $G$ is the Newtonian gravitational constant, and the period $P$
is well known from the photometry. The planet mass $M_p$ is not known
a priori, but its influence is very small compared to the stellar mass
$M_{\star}$, and even a rough value is usually sufficient for this
application. Once the stellar mass is known, the process can be
repeated if necessary with an improved value of $M_p$.

The stellar evolution models we use are those from the Yonsei-Yale
(Y$^2$) series by \cite{Yi:01} \citep[see also][]{Demarque:04}, which
are conveniently provided with tools for interpolating isochrones in
both age and metallicity.\footnote{In addition to the iron abundance,
the enhancement of the $\alpha$ elements can also have a significant
effect on the inferred stellar properties. About half of the
transiting systems have at least one study in the literature reporting
abundances for several of the $\alpha$ elements (Mg, Si, S, Ca, Ti, C,
and O): HD~209458 (3 studies), TrES-1 (2 studies), WASP-1, XO-1, XO-2,
and the five OGLE systems (one study each). In all cases the average
enhancement is not significantly different from zero. We therefore
assume here that it is zero for all systems.}  We explore the full
range of metallicities allowed by the observational errors in [Fe/H],
sampling $\sim$20 equally spaced values for each system. For each
metallicity we consider a range of ages from 0.1 to 14 Gyr, in steps
of 0.1 Gyr. These isochrones are interpolated to a fine grid in mass,
and compared point by point with the measured values of $T_{\rm eff}$
and $a/R_{\star}$. All locations (``matches'') on the isochrone that
are consistent with these quantities within the observational errors
are recorded. We also record the corresponding likelihood given by
$\exp(-\chi^2/2)$, where
\begin{displaymath}
\chi^2 = \left({\Delta{\rm [Fe/H]}\over \sigma_{\rm [Fe/H]}}\right)^2 +
\left({\Delta T_{\rm eff}\over \sigma_{T_{\rm eff}}}\right)^2 +
\left({\Delta(a/R_{\star})\over \sigma_{a/R_{\star}}}\right)^2
\end{displaymath}
and the $\Delta$ quantities represent the difference between the
observed and model values at each point. Observational errors are
assumed to be Gaussian, and the asymmetric error bars in $a/R_{\star}$
were taken into account. The best-fit values for each stellar property
are obtained by computing the sum over all matches, weighted by their
corresponding likelihood.  Additionally, we account for the varying
density of stars on each isochrone prescribed by the Initial Mass
Function (IMF), by multiplying the weights by the number density of
stars at each location as provided with the Y$^2$ isochrones. The IMF
adopted is a power law with a Salpeter index. The effect of this
latter weighting is generally small.

The results for each system are listed in Table~\ref{tab:stellar},
where we give in addition to the mass and radius the theoretical
values of $\log g_{\star}$, luminosity $L_{\star}$, absolute visual
magnitude $M_V$, and evolutionary age. In several cases our results
differ slightly from those reported in recent discovery papers that
use the same atmospheric parameters adopted here and apply the
$a/R_{\star}$ constraint essentially in the same way we have. The
present study represents a slight improvement for those systems due to
the application of weights, as described above.

\begin{figure}
\vskip -0.2in
\epsscale{1.2}
{\hskip -0.1in\plotone{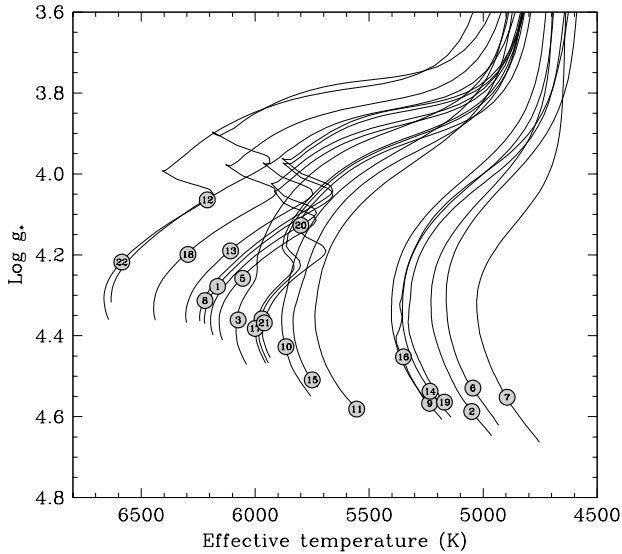}}
\vskip -0.3in

\figcaption[]{Evolutionary tracks for all host stars except GJ~436
from the Yonsei-Yale models of \cite{Yi:01}. The numbering of the
systems follows that in the tables. The masses and metallicities
adopted for the tracks, as well as the location of each star on the
diagram, are from the best fit to the observations.
\label{fig:evolution}}

\end{figure}

\begin{figure}
\vskip -0.2in
\epsscale{1.25}
{\hskip -0.1in\plotone{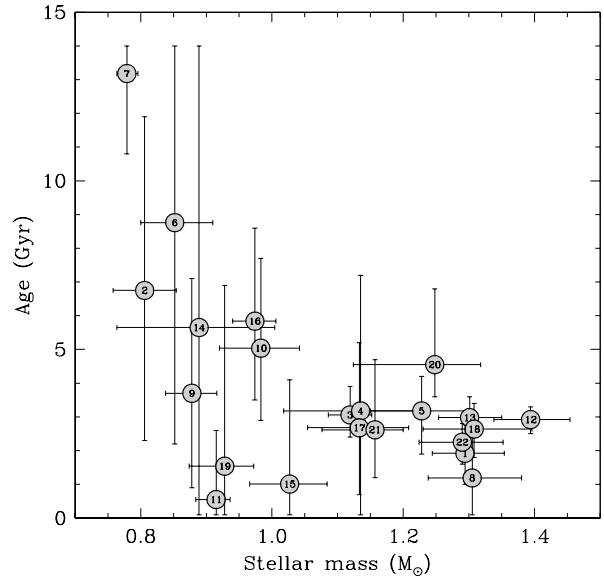}}
\vskip -0.3in

\figcaption[]{Evolutionary age versus mass for transiting planet host
stars based on the models by \cite{Yi:01}. GJ~436 is excluded (see
text). The numbering of the systems is the same as in the
tables.\label{fig:ages}}

\end{figure}

The relative errors of the stellar masses and radii determined here
have median values of about 6\% and 4\%, respectively, but with wide
ranges depending on the precision of the observables (2\% to 13\% for
$\sigma_{M_{\star}}/M_{\star}$, and 1.3\% to 12\% for
$\sigma_{R_{\star}}/R_{\star}$). While all of the stars in the current
sample are hydrogen-burning stars, the degree of evolution within the
main sequence varies considerably (see Figure~\ref{fig:evolution}).
Some systems such as TrES-3 (\#11) and XO-1 (\#15) are near the
zero-age main sequence; others like HAT-P-4 (\#20) have already lived
for $\sim$90\% of their main-sequence lifetime. The age is a critical
ingredient for the theoretical modeling of the structure and evolution
of exoplanets \citep[see, e.g.,][]{Burrows:07}, yet it is among the
most difficult properties to determine for isolated main-sequence
stars. For the host stars of transiting planets the evolutionary ages
span the full range, as seen in Figure~\ref{fig:ages}.  The difficulty
mentioned above is most evident for the less evolved objects
($M_{\star}$ less than about 0.9~M$_{\sun}$), which are all seen to
have large error bars that can reach the upper limit of 14~Gyr
considered here. For those cases the nominal ages reported in this
work should be used with caution.

\begin{figure}
\vskip -0.4in
\epsscale{1.3}
{\hskip -0.2in\plotone{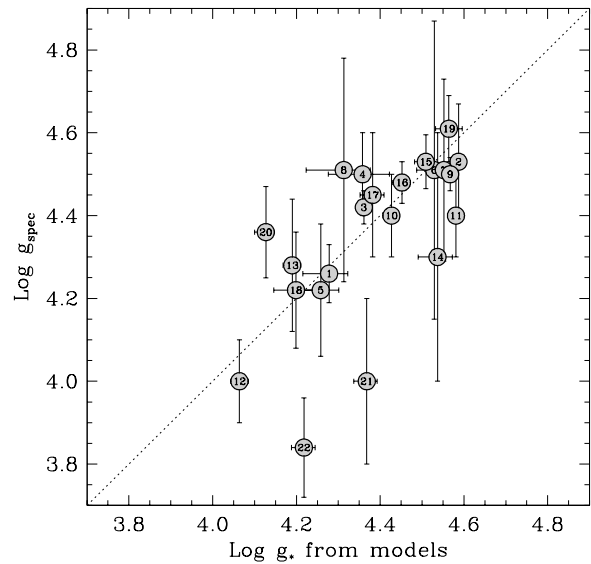}}
\vskip -0.35in

\figcaption[]{Surface gravities for the planet host stars inferred
from stellar evolution models by \cite{Yi:01}, compared with those
measured spectroscopically. The dotted line represents the one-to-one
relation, and the systems are numbered as in previous
figures.\label{fig:logg}}

\end{figure}

\begin{figure}
\vskip -0.2in
\epsscale{1.2}
{\hskip -0.1in\plotone{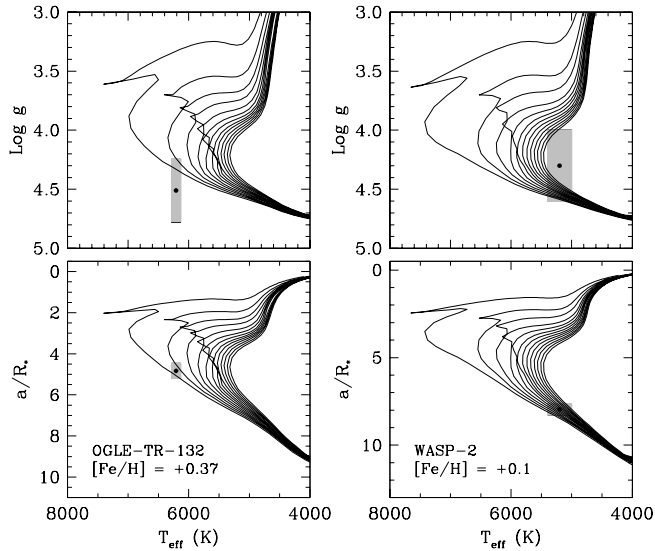}}
\vskip -0.3in

\figcaption[]{Measured properties of two extrasolar planet host stars
displayed on plots analogous to the H-R diagram. The constraint on the
location of the stars based on the surface gravities ($\log g_{\rm
spec}$) and temperatures is shown in the top panels, and the same for
$a/R_{\star}$ versus $T_{\rm eff}$ is shown in the bottom
panels. Isochrones are from the series of evolutionary models by
\cite{Yi:01} for the measured metallicity in each case
(Table~\ref{tab:atmospheric}), and are shown for ages of 1 to 13 Gyr
(left to right) in steps of 1 Gyr.  The $a/R_{\star}$ values are seen
to provide a much better handle on the stellar parameters (mass,
radius, etc.).\label{fig:constraints}}

\end{figure}

The surface gravities inferred from the models ($\log g_{\star}$;
Table~\ref{tab:stellar}) have formal uncertainties that are typically
about 5 times smaller than those measured spectroscopically ($\log
g_{\rm spec}$; Table~\ref{tab:atmospheric}). This reflects the
strength of the constraint provided by $a/R_{\star}$. The values of
$\log g_{\rm spec}$ and $\log g_{\star}$ are compared against each
other in Figure~\ref{fig:logg}. On average they agree quite well (the
mean $O\!-\!C$ difference is $-0.027$ dex), and the rms scatter of the
differences is 0.15 dex although three of the systems present
differences larger than 0.2 dex. An illustration of the superior
constraint afforded by $a/R_{\star}$ is shown in
Figure~\ref{fig:constraints} for OGLE-TR-132 and WASP-2, two of the
more dramatic examples. In neither case is the parallax known.

\section{Comparison with other models}
\label{sec:modchecks}

\begin{figure}
\vskip 0.3in
\epsscale{1.7}
{\hskip -0.7in\plotone{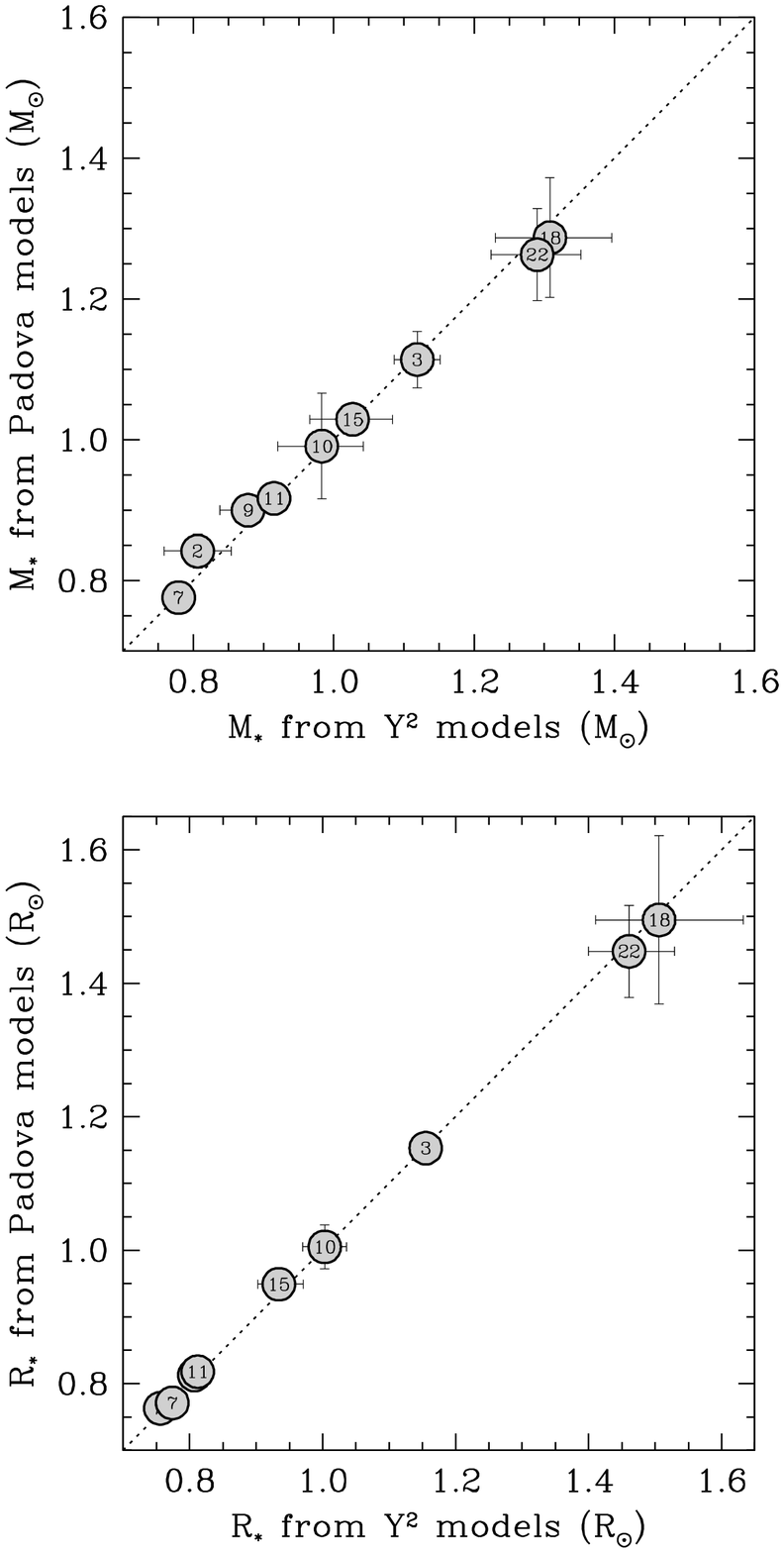}}
\vskip 0.75in

\figcaption[]{Comparison of the stellar masses and radii derived from
the Y$^2$ models \citep{Yi:01} and the Padova models
\citep{Girardi:00} for transiting planet hosts, showing the excellent
agreement. The dotted lines represent the one-to-one relation, and the
numbering of the systems is the same as in previous figures.
\label{fig:checks}}

\end{figure}

There is undoubtedly some systematic error introduced by imperfections
in the stellar evolution models themselves, but this type of error is
difficult to evaluate. Extensive comparisons between models and
observations using double-lined eclipsing binaries with very
accurately measured masses and radii have shown that the agreement
with theory is in general very good, and is within a few percent for
solar-type stars \citep[see, e.g.,][]{Andersen:91, Pols:97,
Lastennet:02}.  As a simple test, we considered a second set of models
by \cite{Girardi:00} that has often been used by other investigators
in the field of transiting planets.  We used the isochrone
interpolation tools provided on the web site at the Osservatorio
Astronomico di Padova\footnote{\tt
http://stev.oapd.inaf.it/$\sim$lgirardi/cgi-bin/cmd.} to explore the
agreement with the observed values of $T_{\rm eff}$, [Fe/H], and
$a/R_{\star}$, and to infer the stellar mass and other properties in
the same way as above. There are small differences in the physical
assumptions between these models and those from the Y$^2$ series, but
overall they are rather similar. For this application the Padova
models are available for metallicities smaller than $Z = 0.030$
(corresponding to [Fe/H] $= +0.20$), which allows us to compare
results for nine of the transiting
systems.\footnote{Higher-metallicity Padova models have been published
by \cite{Salasnich:00}, but those employ physical assumptions
different enough that they cannot be merged with the \cite{Girardi:00}
models we use here. In addition, from a practical point of view, the
use of these higher-metallicity models is not yet implemented in the
online interpolation routine as of this writing.}
Figure~\ref{fig:checks} displays this comparison for the stellar
masses and radii, showing that in all cases the results from both
models agree to well within their uncertainties.

Similar tests were carried out using the models of \cite{Baraffe:98}.
These models enjoy widespread use for lower-mass stars and brown
dwarfs, but they are also computed for masses as large as
1.4~M$_{\sun}$. They are available for three different values of the
mixing length parameter $\alpha_{\rm ML}$ over restricted ranges in
mass and metallicity.  The value that fits the observed properties of
the Sun (against which all models are calibrated) is $\alpha_{\rm ML}
= 1.9$. For cooler objects near the bottom of the main sequence the
value used almost universally is $\alpha_{\rm ML} = 1.0$.  Here we
adopt the former value, since our stars are all more massive than
about 0.8~M$_{\sun}$ with the exception of GJ~436, which we discuss in
more detail in \S\,\ref{sec:gj436}. For the mass range of interest and
for $\alpha_{\rm ML} = 1.9$ the \cite{Baraffe:98} models are publicly
available only for solar metallicity, so the comparison was limited to
transiting systems with [Fe/H] within about 0.1 dex of the Sun, and
with [Fe/H] uncertainties that keep them within 0.2 dex of solar. Of
the six systems in this range, one (OGLE-TR-113) gave no solution
consistent with the observed values of $T_{\rm eff}$ and
$a/R_{\star}$. This is because the Baraffe models are limited to ages
less than about 10~Gyr for this mass range, whereas both the
\cite{Yi:01} and the \cite{Girardi:00} models indicate an age of about
13 Gyr for this star. Whether OGLE-TR-113 is truly this old is
unknown. It seems at least as likely that some of the other
observational constraints are in error. The remaining 5 systems show
excellent agreement with the results from both the Y$^2$ models and
the Padova models, within the formal uncertainties.

These tests of the stellar evolutionary models are obviously not
exhaustive, and it may well be the case that all of the sets of models
we considered have some deficiencies in common. However, the general
pattern of agreement does lend some degree of confidence to the
results.  We proceed under the assumption that the systematic errors
in these calculations are not the dominant source of error in the
stellar parameters derived here (except perhaps for OGLE-TR-113, as
noted above).

\section{Additional observational constraints}
\label{sec:obschecks}

As a further test of the accuracy of our stellar radius
determinations, in this section we consider the consistency check
provided by the near infrared (NIR) surface brightness (SB) relations,
which yield the angular diameter $\phi$ of a star directly in terms of
its apparent magnitude and color. Unlike the parallax, which would in
principle yield the stellar luminosity but is known for only five of
the brighter systems in the sample, $\phi$ can be computed for all
host stars from existing photometry. We use the empirical relation
\begin{equation}
\label{eq:kervella}
\log \phi_{\rm SB} = c_1 (V\!-\!K) + c_2 - 0.2 K
\end{equation}
derived by \cite{Kervella:04}, in which the coefficients are $c_1 =
0.0755 \pm 0.0008$ and $c_2 = 0.5170 \pm 0.0017$, the $K$-band
magnitude is in the Johnson system, and $\phi_{\rm SB}$ is the
limb-darkened value of the angular diameter expressed in milli-arc
seconds. The relation represented by eq.\,(\ref{eq:kervella}) is
extremely tight, with a scatter well under 1\%. Near infrared
magnitudes are available for all stars from the 2MASS catalog and were
transformed to the Johnson system following \cite{Carpenter:01}.  The
best available $V$ magnitudes collected from the literature are listed
in Table~\ref{tab:atmospheric}.

Angular diameters from our stellar evolution modeling in
\S\,\ref{sec:masses} can be derived for comparison with $\phi_{\rm
SB}$ by making use of our theoretical radii and absolute visual
magnitudes in Table~\ref{tab:stellar}, along with the apparent $V$
magnitudes, using
\begin{equation}
\label{eq:modelphi}
\phi_{\rm mod} = 9.3047 R_{\star}/10^{0.2(V-M_V+5)}.
\end{equation}
With the stellar radius expressed in solar units, the numerical
constant is such that $\phi_{\rm mod}$ is in mas. Neither this
equation nor the previous one take into account interstellar
extinction, although eq.\,(\ref{eq:kervella}) is actually quite
insensitive to extinction. We discuss this below.

\begin{figure}
\vskip -0.55in
\epsscale{1.3}
{\hskip -0.1in\plotone{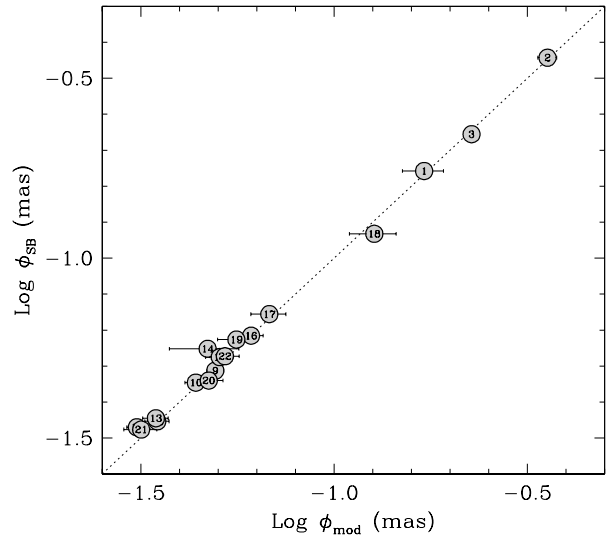}}
\vskip -0.3in

\figcaption[]{Angular diameters $\phi_{\rm SB}$ computed from the near
infrared surface brightness relation of \cite{Kervella:04} compared
against the values derived from our modeling, making use of the
apparent $V$ magnitudes for all host stars and ignoring
extinction. The dotted line represents the one-to-one relation, and
the systems are numbered as in previous figures.\label{fig:phi}}

\end{figure}

The comparison between the angular diameters from
eq.\,(\ref{eq:kervella}) and eq.\,(\ref{eq:modelphi}) is shown in
Table~\ref{tab:phi} for all systems except for GJ~436, for reasons to
be described in the next section, and the OGLE stars, which are likely
to be significantly affected by extinction since they lie several kpc
away near the Galactic plane.  The uncertainties listed include all
contributions from the photometry and model-derived quantities, as
well as the errors in the coefficients of
eq.\,(\ref{eq:kervella}). The precision of $\phi_{\rm SB}$ is
typically several times better than that of $\phi_{\rm mod}$.  A
graphical comparison is shown in Figure~\ref{fig:phi}, with the
one-to-one relation represented with a diagonal line. The good
agreement over the full range of an order of magnitude in $\phi$ is an
indication that our model radii from \S\,\ref{sec:masses} are
accurate.

One effect of extinction in this analysis would be to make the values
based on eq.\,(\ref{eq:modelphi}) appear too small. Examination of the
differences shows a hint of this, particularly among the fainter (more
distant) stars: the average difference $\left<(\phi_{\rm
mod}-\phi_{\rm SB})/\phi_{\rm SB}\right>$ is $+0.5 \pm 1.8\%$ for the
6 objects brighter than $V = 11$, and $-3.9 \pm 1.6\%$ for the others.
Since reddening values for individual systems are presently unknown,
and cannot be determined accurately enough from the information at
hand, we are unable to correct for this. We are equally unable to turn
the argument around and use the $\phi_{\rm SB}$ values as further
constraints in our modeling. By trial and error we find that a uniform
reddening value between $E(B\!-\!V) = 0.02$ and 0.03 is sufficient to
reduce the average discrepancy in the angular diameters for the
fainter stars to zero. This modest amount of reddening is consistent
with expectations for objects that are between 100 and 500 pc away, as
are these.

In addition to the angular diameters, Table~\ref{tab:phi} lists the
parallaxes $\pi_{\rm mod}$ and corresponding distances we derive from
$V$ and $M_V$, ignoring extinction. \hip\ parallaxes $\pi_{\rm HIP}$
are given for comparison as well, for the objects that have them.

\section{The case of GJ~436}
\label{sec:gj436}

As the only M dwarf among the currently known transiting planet host
stars, GJ~436 presents a special challenge. The Y$^2$ stellar
evolution models we used in all other cases are not intended for the
lower main sequence, as they lack the proper non-grey model atmosphere
boundary conditions to the interior equations that have been shown to
be critical for cool objects \citep[see, e.g.,][]{Chabrier:97}. The
Padova models of \cite{Girardi:00} have the same shortcoming, although
both of these models are perfectly adequate for hotter stars. On the
other hand, the calculations by \cite{Baraffe:98} with $\alpha_{\rm
ML} = 1.0$ are specifically designed for low-mass stars, and have in
fact been invoked by other authors for this system as a rough
consistency check. For the most part, the previous mass determinations
for GJ~436 have relied upon empirical mass-luminosity relations, but
even those relations have their problems. Previous radius estimates
for GJ~436 have often rested on the assumption of numerical equality
between $M_{\star}$ and $R_{\star}$ for M stars \citep{Gillon:07a,
Deming:07}.

The importance of the GJ~436 system is undeniable, as it harbors the
smallest transiting planet that has been found to date (comparable in
size to Neptune). Furthermore, it is the nearest transiting exoplanet,
at a distance of only 10 pc.  Given these facts, it is somewhat
surprising that until recently the mass of the star
($\sim$0.4~M$_{\sun}$) was only known to about 10\% \citep{Maness:07,
Gillon:07b}, on a par with the worst of the determinations in
Table~\ref{tab:stellar} (that of WASP-2, with poorly determined
spectroscopic parameters and more than 10 times farther away). Similar
limitations hold for the stellar radius. As a result, our knowledge of
the planetary parameters has suffered.  \cite{Torres:07b} showed that
the \cite{Baraffe:98} models in their original form do not yield
satisfactory results for GJ~436 since different answers for
$M_{\star}$ and $R_{\star}$ are obtained depending on which
constraints are used (including the absolute magnitude, since a
reliable parallax measurement is available from \hip).
Moreover, some of the other predicted quantities are at odds with
empirical determinations.  These difficulties had not been previously
emphasized.

However, as described by \cite{Torres:07b}, reliable parameters can
still be obtained from the models by modifying the theoretical radii
and temperatures in such a way as to preserve the bolometric
luminosities. This recognizes the fact that $R_{\star}$ and $T_{\rm
eff}$ as predicted by theory have been shown to disagree with accurate
measurements for M dwarfs in double-lined eclipsing binaries
\citep[see, e.g.,][]{Popper:97, Clausen:99, Torres:02, Ribas:03,
Lopez-Morales:05}, whereas the luminosities appear to be unaffected
\citep{Delfosse:00, Torres:02, Ribas:06, Torres:06}. \cite{Torres:07b}
applied simultaneously the observational constraints on $T_{\rm eff}$,
$a/R_{\star}$, the color index $J\!-\!K$, and the absolute $K$-band
magnitude $M_K$, and allowed the radius/temperature adjustment factor
to be a free parameter. In this way a self-consistent solution was
achieved for all parameters, and in addition the radius/temperature
factor showed excellent agreement with previous estimates for other M
dwarfs. The stellar mass and radius are $M_{\star} =
0.452_{-0.012}^{+0.014}$~M$_{\sun}$ and $R_{\star} =
0.464_{-0.011}^{+0.009}$~R$_{\sun}$.  We adopt these results here,
thus placing GJ~436 on a similar footing as the other transiting
systems in that the properties of the host stars are all based on
current stellar evolution models.

The results for GJ~436 are listed in Table~\ref{tab:stellar} along
with those of the other 22 stars. A few words of caution are in
order. As pointed out by \cite{Torres:07b}, the predicted absolute
visual magnitude for this star is unreliable due to missing molecular
opacity sources shortward of 1~$\mu$m in the model atmospheres that
are used as boundary conditions in the stellar evolution calculations
\citep[see, e.g.,][]{Baraffe:98, Delfosse:00}.  Also, because of the
unevolved status of the star, the nominal age is probably meaningless
since the observational constraints allow any age within the range of
1--10 Gyr explored in the modeling effort of \cite{Torres:07b}.
Finally, this modeling effort assumed that the stellar metallicity is
solar. For GJ~436 this is a valid assumption since the measured
composition is [Fe/H] $= -0.03 \pm 0.20$ (see Appendix), but the
uncertainty in [Fe/H] was not accounted for due to the lack of proper
models, and therefore the errors of the stellar properties in
Table~\ref{tab:stellar} may be underestimated.

A careful reader may wonder whether our use of the \cite{Baraffe:98}
models in \S\,\ref{sec:modchecks} as a check on the results from the
Y$^2$ isochrones is contradicted by our remarks on the
radius/temperature discrepancies mentioned above. We do not believe
so, because the mass regime of GJ~436 is very different. The host
stars of the other transiting planets are typically more than twice as
massive as GJ~436.

\section{Planetary parameters}
\label{sec:results}

The combination of the stellar properties in
Table~\ref{tab:atmospheric} with the light curve parameters in
Table~\ref{tab:lcfits}, along with the measured orbital periods and
velocity semi-amplitudes, yields the homogeneous set of planet
parameters presented in Table~\ref{tab:planetary}.  In addition to the
planetary mass and radius, we have computed and tabulated the
planetary surface gravity and mean density, as well as the orbital
semimajor axis and other useful characteristics. As pointed out by
\cite{Southworth:07}, Winn et al.~(2007), \cite{Beatty:07}, and
others, the surface gravity of a transiting planet can be derived
without knowledge of the stellar mass or radius, using only the
radial-velocity curve and the transit light curve.  We take this
opportunity to correct the general expression for $\log g_p$ presented
by \cite{Sozzetti:07}, valid also for the case of eccentric orbits,
which neglected to account for the projection factor implicit in the
impact parameter $b$ as derived from the light-curve fits:
\begin{eqnarray}
\label{eq:gravity}
\log g_p & = & -2.1383 - \log P + \log K_\star - \\ \nonumber
& - & {1\over 2}\log \left(1-\left[{b \over a/R_\star}
{1-e^2\over 1+e\sin\omega}\right]^2\right) + \\ \nonumber
& + & 2 \log\left({a/R_\star \over R_p/R_\star}\right) +
{1\over 2}\log (1-e^2)~.
\end{eqnarray}
The numerical constant is such that the gravity is in {\it cgs\/}
units when $P$ and $K_\star$ are expressed in their customary units of
days and m~s$^{-1}$.

\begin{figure}      % bitmapped figure
\vskip -0.01in
\epsscale{1.1}
{\hskip 0.1in\plotone{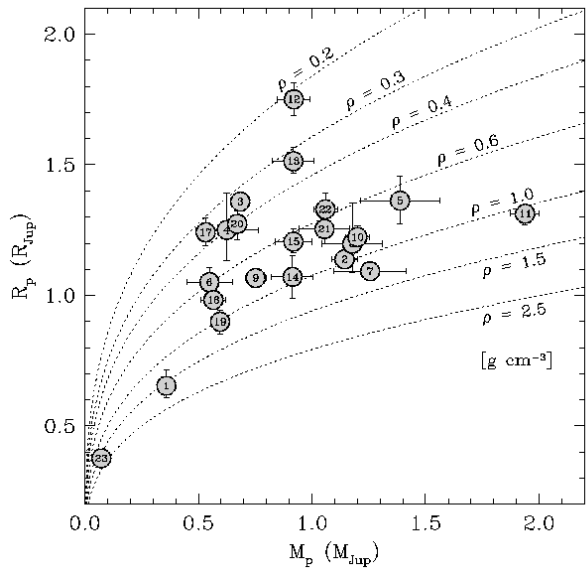}}
\vskip -0.05in

\figcaption[]{Mass-radius diagram for all transiting planets (HAT-P-2
is off the scale, with $M_p = 8.7$~M$_{\rm Jup}$). Lines of constant
density are shown.\label{fig:rm}}

\end{figure}

The first properties one generally wants to know about a transiting
planet are its mass and radius. The large size measured for the first
transiting planet discovered, HD~209458b, has been widely debated and
still presents somewhat of a challenge to theory. It is now
accompanied by several other inflated planets, underlying our
incomplete knowledge of the physics of these objects.  An updated
version of the now classical diagram of $M_p$ versus $R_p$ is shown in
Figure~\ref{fig:rm}, in which five other examples are seen to be at
least as large as HD~209458b (\#3), or perhaps even larger, given the
uncertainties.

%\begin{figure}     % original figure
%\vskip -0.4in
%\epsscale{1.3}
%{\hskip -0.2in\plotone{f7.eps}}
%\vskip -0.3in

%\begin{figure*}    % original figure
%\vskip -0.3in
%\epsscale{1.20}
%\plotone{f8.eps}
%\vskip 1.15in

\begin{figure*}    % bitmapped figure
\epsscale{1.2}
\vskip -0.1in
\plotone{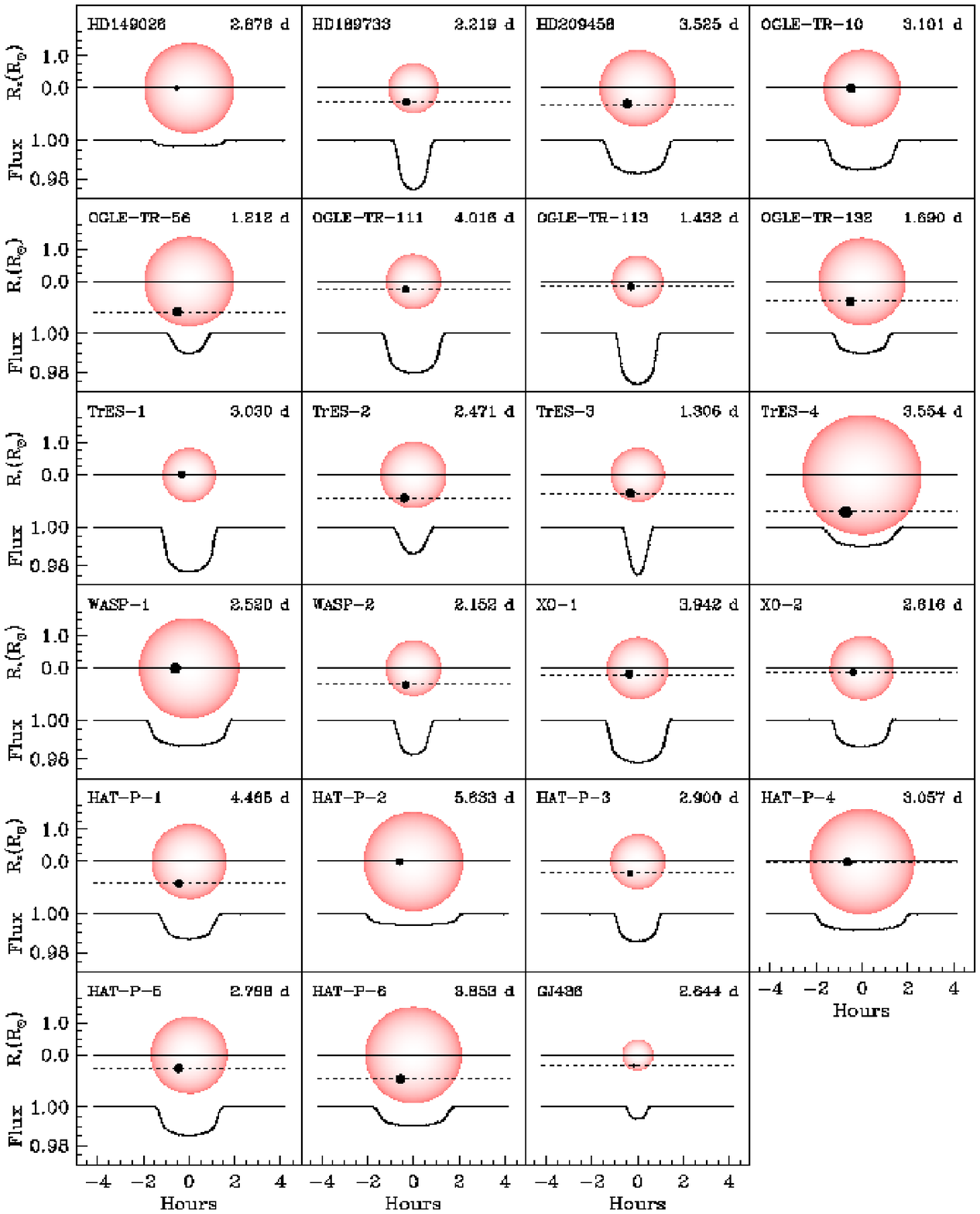}
\vskip 0.05in

\figcaption[]{``Portrait gallery'' of transiting extrasolar
planets. Star and planet sizes are shown to scale.  The vertical and
horizontal axes of all light curves are also on the same scale. Planet
trajectories are shown with their measured impact parameters (dotted
lines). Orbital periods are indicated in each
panel.\label{fig:gallery}}

\end{figure*}

The fundamental physical properties of the known transiting planets
cover a considerable range---more than two orders of magnitude in
mass, a factor of nearly 5 in radius, and a factor of 3 in orbital
separation---and these are discussed in the next section. But the
geometric properties that determine the light-curve shapes are also
varied. In Figure~\ref{fig:gallery} we present a ``portrait gallery''
of all known transiting planets. Stars and planets are rendered to
scale, emphasizing the wide range of stellar types probed by the
photometric searches. The orbital geometries are indicated with solid
horizontal lines representing an edge-on orientation, and the path of
each planet across the stellar disk shown with dotted lines at the
actual impact parameter. (Of course, the data do not distinguish
between positive and negative impact parameters.) The resulting light
curves calculated for the $V$ band are all shown with the same
vertical (flux) and horizontal (time) scale, to facilitate comparison.
Depths vary quite significantly (3~mmag in $V$ for HD~149026b, 26~mmag
for OGLE-TR-113b; see Table~\ref{tab:lcfits}), as do the overall
shapes of the transit events, which in several cases are grazing
enough to depart from the canonical profiles often depicted in the
literature.

\section{Discussion}
\label{sec:discussion}

Many investigators have sought and claimed possible correlations
between various stellar and planetary parameters of transiting
systems.  In principle such correlations could lead to important
insights into the formation, structure, and evolution of exoplanets.
A primary motivation for presenting a more complete, accurate, and
homogeneous set of these parameters in this work was to facilitate
such studies. The relatively large array of properties now available
offers the opportunity to find new correlations, or to revisit old
ones incorporating additional variables. While it is beyond the scope
of the present work to investigate all possible correlations with
statistical rigor, in this section we check on three of the most
intriguing and potentially important relations that have been
proposed.

\subsection{Planetary mass versus orbital period}
\label{sec:massperiod}

\begin{figure}        % bitmapped figure
\vskip 0.0in
\epsscale{1.3}
{\hskip -0.2in\plotone{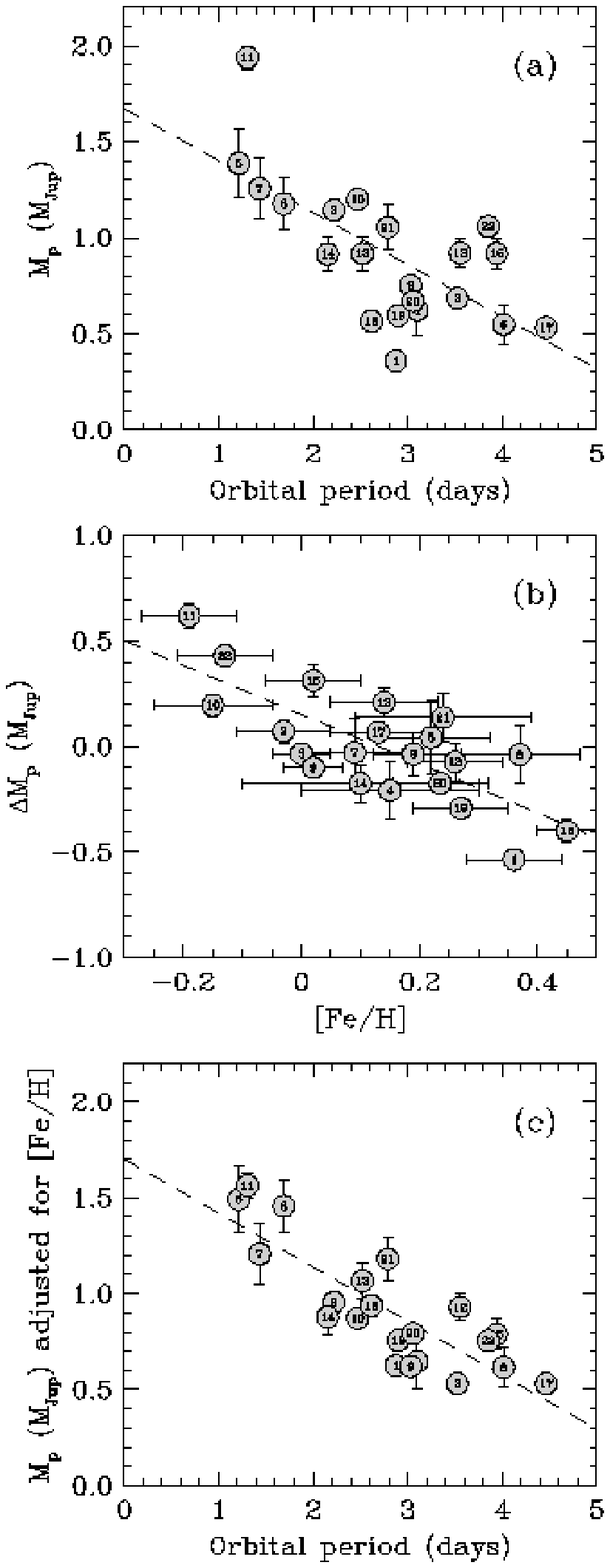}}
\vskip 0.0in

\figcaption[]{(a) $M_p$ as a function of period for all transiting
planets except HAT-P-2b ($M_p = 8.7$~M$_{\rm Jup}$, $P = 5.63$~d) and
GJ~436b ($M_p = 0.073$~M$_{\rm Jup}$, $P = 2.64$~d); see text. The
dashed line is a linear fit. (b) $O\!-\!C$ residuals $\Delta M_p$ from
the linear fit in the top panel shown as a function of the metallicity
of the host star. The dashed line is a linear fit. (c) Same as (a),
with the dependence on [Fe/H] removed based on the fit in (b). The
fitted line has the expression $M_p = (+1.70 \pm 0.13) - (0.281 \pm
0.044) \times P$.
\label{fig:mp}}

\end{figure}

\cite{Mazeh:05} were the first to point out the apparent correlation
between $M_p$ and $P$ for transiting planets \citep[see
also][]{Gaudi:05}. The original suggestion was based on only 6
systems, but additional discoveries have generally supported the trend
of decreasing mass with longer periods, although the scatter has also
become larger. This is shown in Figure~\ref{fig:mp}a. HAT-P-2b would
be an extreme outlier in this plot; we have excluded it because it is
so much more massive than the other planets and may belong to a
different category of planet (see \S\,\ref{sec:safronov}). Similarly,
we have excluded GJ~436b because it is so much less massive than the
others, and may be a rocky or rock-ice planet rather than a gas
giant. A simple linear fit is shown for reference (dashed line).

%\begin{figure}        % original figure
%\vskip 0.35in
%\epsscale{2.4}
%{\hskip -1.9in\plotone{f9.eps}}
%\vskip 1.2in

We also investigated the scatter in this relation, seeking any ``third
variable'' that might correlate with the residuals.  We seem to have
found such a third variable: the metallicity of the host star.
Figure~\ref{fig:mp}b displays the $O\!-\!C$ residuals from the top
panel as a function of [Fe/H], indicating a rather clear correlation
(dashed line): $\Delta M_p = (+0.152 \pm 0.050) - (1.17 \pm 0.23)
\times {\rm [Fe/H]}$. After removal of this trend, the relation
between $M_p$ and period becomes tighter (Figure~\ref{fig:mp}c).  The
scatter in the mass-period relation is reduced from 0.26~M$_{\rm Jup}$
in the top panel to 0.17~M$_{\rm Jup}$ in the bottom panel. It seems
unlikely that this is a statistical fluke. However, the scatter is
still larger than the formal observational uncertainties, suggesting
that these three variables are not completely determinative.

What might be the implications of this metallicity dependence?  It has
been proposed that the mass-period relation is related to the process
by which close-in exoplanets migrated inward from their formation
sites, or more specifically, to the mechanism that halts migration at
orbital periods of a few days. The trend of larger masses at shorter
orbital periods could suggest that the halting mechanism depends on
mass, and larger planets are able to migrate further in. The
dependence on metallicity may then be interpreted to indicate that
planets in metal-poor systems need to be more massive in order to
migrate inward to the same orbital period as more metal-rich
planets. In this context, the trend in Figure~\ref{fig:mp}b could be
interpreted as evidence that the efficiency of the migration (or
halting) mechanism is affected to some degree by the chemical
composition.  A dependence of migration on metallicity is in fact
predicted by some theories, and could arise, as pointed out by
\cite{Sozzetti:06}, either from slower migration rates in metal-poor
protoplanetary disks \citep{Livio:03, Boss:05} or through longer
timescales for giant planet formation around metal-poor stars, which
would effectively reduce the efficiency of migration before the disk
dissipates \citep{Ida:04, Alibert:05}. However, the above processes
are more aimed at addressing the apparent lack of short-period planets
among very metal-poor stars claimed by some authors
\citep[see][]{Sozzetti:06}, whereas among the transiting planets it is
not the lack of more metal-poor examples we are concerned with, but
rather their different properties (such as mass) compared to
metal-rich planets \emph{at the same orbital period}.  To give a
quantitative example, we find that for a period of 2.5 days (near the
average for known transiting systems), the mass of a planet with
[Fe/H] $= -0.2$ is $\sim$40\% larger than one with average metallicity
([Fe/H] = +0.13, excluding HAT-P-2b and GJ~436b), while the mass of a
planet with [Fe/H] $= +0.4$ is about 30\% smaller. This range of
metallicities, from [Fe/H] $= -0.2$ to +0.4 (covering a factor of 4 in
metal enhancement), is approximately the full range observed. We note
that the strength of the metallicity effect is modest rather than
overwhelming, and this too is in agreement with theoretical
expectations \citep[e.g.,][]{Livio:03}. The massive planet HAT-P-2b
obviously does not conform to this trend of $M_p$ versus period, which
may indicate some fundamental difference either in its formation or
migration.

An alternative interpretation, also proposed by \cite{Mazeh:05}, is
that the $M_p$ versus $P$ relation is more a reflection of survival
requirements in close proximity to the star, due to thermal
evaporation from the extreme UV flux. Close-in planets must be more
massive to avoid ablation to the point of undetectability. The role of
metallicity in this case would be through the difference in the
internal structure \citep{Santos:06}. Metal-rich planets have been
suggested to be more likely to develop rocky cores \citep[][see also
\S\,\ref{sec:cores}]{Pollack:96, Guillot:06, Burrows:07}. If the
presence of such a core somehow slows down or prevents complete
evaporation, as has been proposed \citep{Baraffe:04, Lecavelier:04},
survival at a given period would then have a dependence on
metallicity.

\begin{figure}
\vskip -0.35in
\epsscale{1.3}
{\hskip -0.2in\plotone{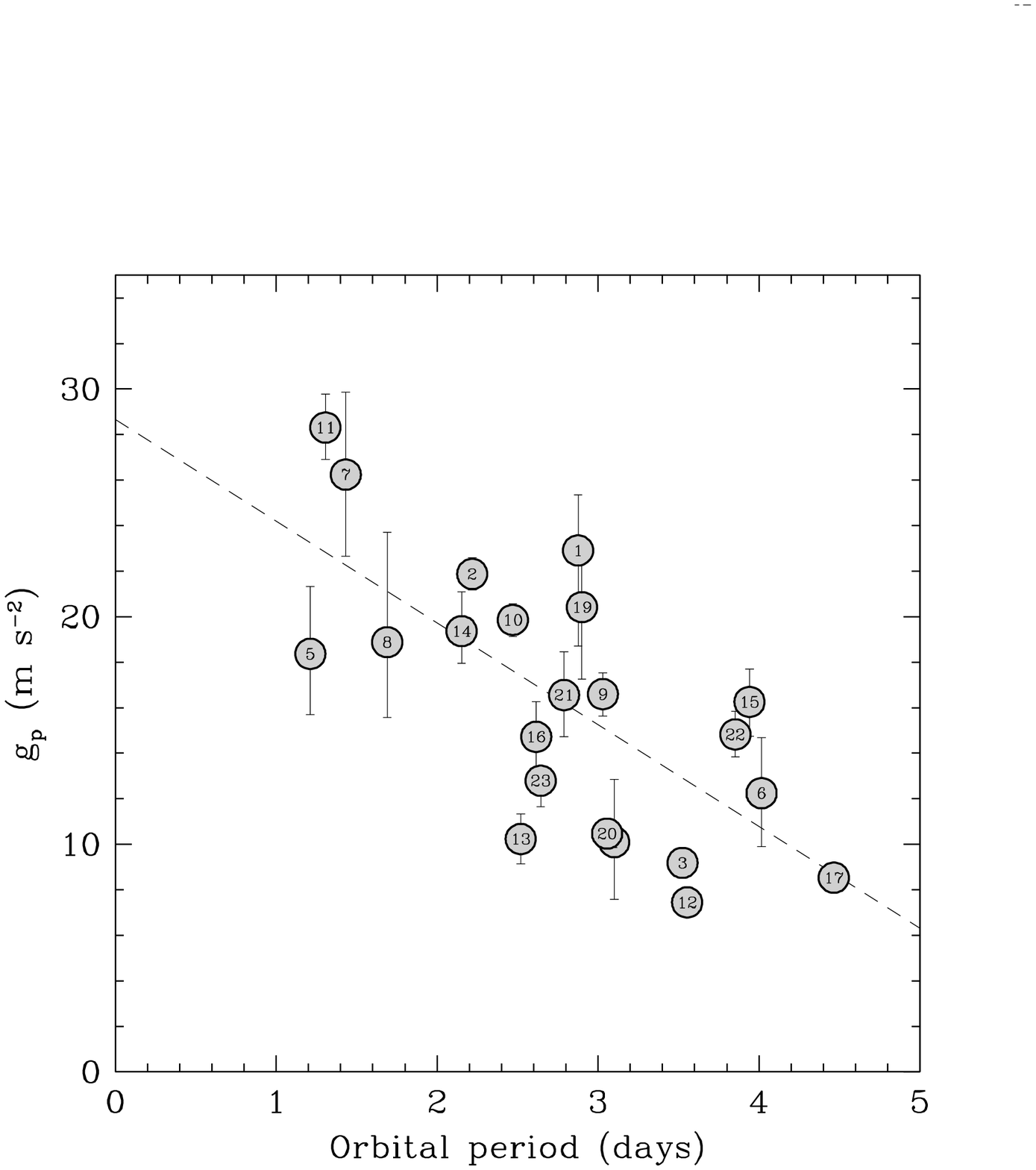}}
\vskip -0.3in

\figcaption[]{Surface gravity versus orbital period for all transiting
planets except HAT-P-2, which is off the scale ($g_p \sim 234$
m~s$^{-2}$). A linear fit is shown.\label{fig:loggperiod}}

\end{figure}

A diagram related to the one considered above is that of planetary
surface gravity versus orbital period, first presented by
\cite{Southworth:07}.  Because $g_p$ does not require knowledge of the
mass or radius of the star, it is in a sense a cleaner quantity that
should be free from systematic errors in $M_{\star}$ and $R_{\star}$
and is nearly independent of stellar evolution models. This not true
of other bulk properties such as the mean planet density. An updated
version of the $g_p$ versus $P$ relation is shown in
Figure~\ref{fig:loggperiod}, along with a linear fit.  We examined the
residuals from this linear fit for a possible correlation with stellar
metallicity, as we did earlier for the case of the $M_p$ versus $P$
diagram, but we found none.  Given that $g_p \propto M_p/R_p^2$, we
note that the apparent influence of metallicity on the planetary radii
must be contributing significantly to the scatter in the $g_p$ versus
period relation. We discuss this further in \S\,\ref{sec:cores}.

\subsection{Safronov number versus equilibrium temperature}
\label{sec:safronov}

Recently, \cite{Hansen:07} proposed a distinction between two classes
of hot Jupiters, based on a consideration of the Safronov number and
the zero-albedo equilibrium temperature. The Safronov number is a
measure of the ability of a planet to gravitationally scatter other
bodies \citep{Safronov:72}, and is defined as $\Theta =
\frac{1}{2}(V_{\rm esc}/V_{\rm orb})^2 = (a/ R_p)(M_p/ M_{\star})$,
the ratio between the escape velocity and the orbital velocity
squared.  We assume that the zero-albedo equilibrium temperature
scales as $T_{\rm eq} = T_{\rm eff}(R_{\star}/2a)^{1/2}$ (i.e., we
assume that the heat redistribution factor $f$ is common to all
planets, in the absence of more complete knowledge).

We list $\Theta$ and $T_{\rm eq}$ for all transiting planets in
Table~\ref{tab:planetary}.  \cite{Hansen:07} pointed out a gap in the
distribution of Safronov numbers, and defined Class~I planets as those
with $\Theta \sim 0.07 \pm 0.01$ and Class~II as $\Theta \sim 0.04 \pm
0.01$.  They tentatively proposed also that these two categories have
other distinguishing characteristics, such as a difference in the
average temperature of the host stars, or the orbital
separations. Upon the discovery of HAT-P-5b \citep{Bakos:07c} and
HAT-P-6b \citep{Noyes:07}, those authors pointed out that the
distinction now seems less clear, as these two planets tend to fill
the gap between Class~I and Class~II.

\begin{figure}
\vskip -0.35in
\epsscale{1.3}
{\hskip -0.2in\plotone{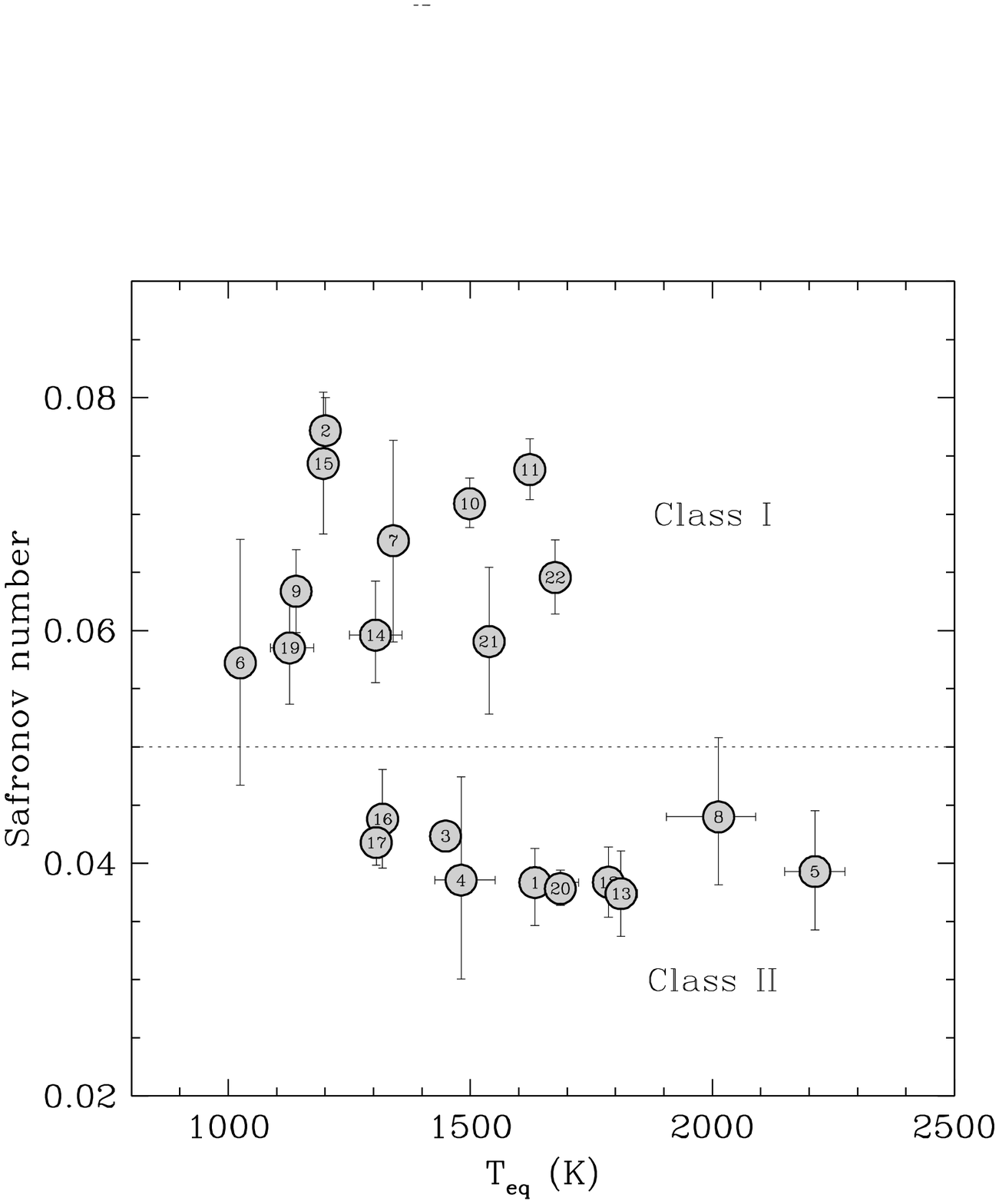}}
\vskip -0.3in

\figcaption[]{Diagram of Safronov number ($\Theta$) versus equilibrium
temperature ($T_{\rm eq}$) for transiting planets. Planet classes are
labeled following \cite{Hansen:07}. A tentative dividing line at
$\Theta = 0.05$ is indicated. \label{fig:safronov}}

\end{figure}

The larger sample now available, and especially the more accurate and
homogeneous set of properties presented here, offers the opportunity
to revisit the issue. Figure~\ref{fig:safronov} shows an updated
version of the $\Theta$--$T_{\rm eq}$ diagram for transiting planets,
which has some significant differences compared to the original
version. Following \cite{Hansen:07} we have excluded the massive
planet HAT-P-2b, with a Safronov number so much larger than all the
others ($\Theta = 0.94$) that it would seem to be in a different class
altogether, as well as GJ~436b, which is of much lower mass and a
presumably different composition.

\begin{figure}
\vskip -0.35in
\epsscale{1.3}
{\hskip -0.2in\plotone{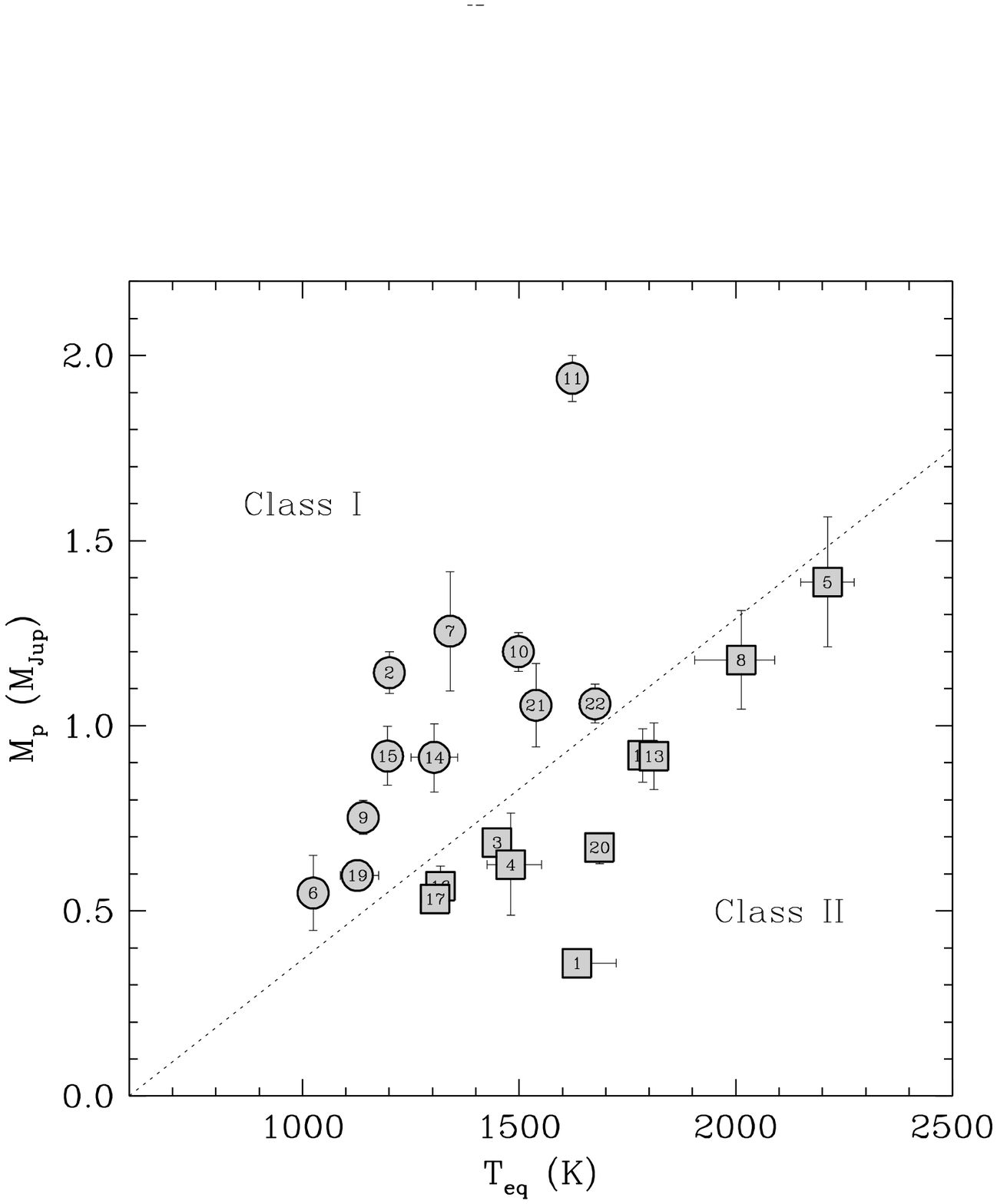}}
\vskip -0.3in

\figcaption[]{Planetary mass as a function of equilibrium temperature
for transiting planets. Class~II planets ($\Theta < 0.05$) are
represented with squares. A tentative dividing line is indicated.
\label{fig:masstemp}}

\end{figure}

In our updated diagram the separation between Class~I and Class~II is
still quite striking. The clustering of the Class~II objects around
the value $\Theta \sim 0.04$ has tightened, if anything, and HAT-P-5b
(\#21) and HAT-P-6b (\#22) do not encroach on the gap in Safronov
numbers with Class~I. (The Class~II object with the lowest value of
$\Theta$ is OGLE-TR-111b [\#6].) Thus, the dichotomy remains very
suggestive.  \cite{Hansen:07} have argued that the principal
distinction between the two classes is based on mass (planetary and/or
stellar), and that planets of Class~II are, on average, less massive
than those in Class~I, and orbit stars that are typically more
massive.  While we agree with the latter part of this statement
regarding the \emph{stellar} masses, we find that the distribution of
\emph{planetary} masses is indistinguishable between the two
groups. We do confirm, however, that an updated diagram of $M_p$
versus $T_{\rm eq}$ (see Figure~\ref{fig:masstemp}) shows planets in
Class~II to be systematically less massive \emph{for the same
equilibrium temperature} (level of irradiation), as was also found by
\cite{Hansen:07}, so in this sense their general claim that the
distinction has something to do with mass seems to be supported by the
observations.

\begin{figure}
\vskip -0.3in
\epsscale{1.3}
{\hskip -0.2in\plotone{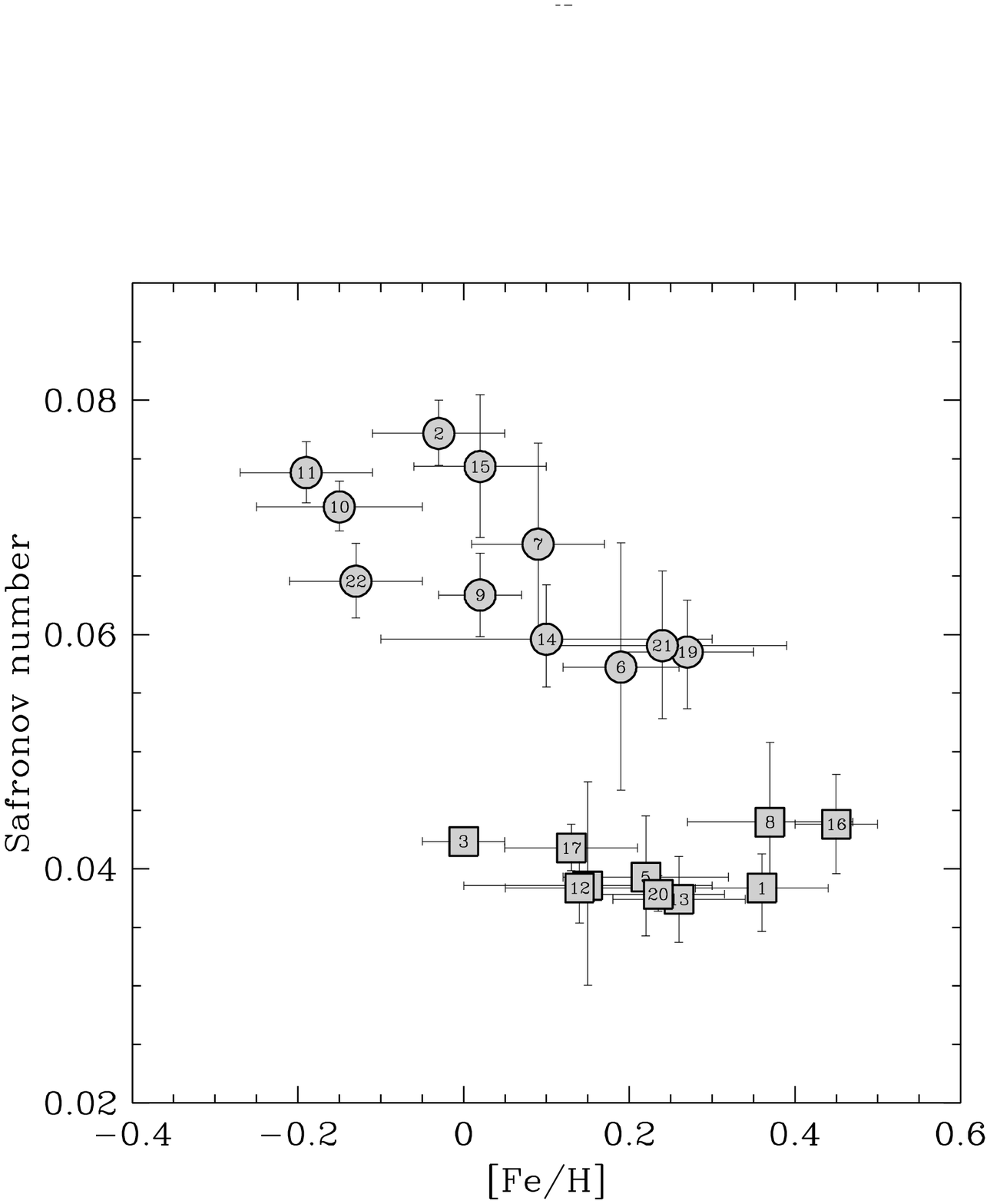}}
\vskip -0.3in

\figcaption[]{Safronov number as a function of stellar metallicity for
transiting planets. Class~II planets ($\Theta < 0.05$) are represented
with squares.\label{fig:saffeh}}

\end{figure}

An important characteristic that seems to be different in the two
groups is the metallicity of the parent stars. This is illustrated in
Figure~\ref{fig:saffeh}. Parent stars of Class~II planets tend to be
slightly more metal-rich. A Kolmogorov-Smirnov test of the two
metallicity distributions, which appear to be centered around [Fe/H]
$\sim 0.0$ for Class~I and [Fe/H] $\sim +0.2$ for Class~II, indicates
only a 1.7\% probability that they are drawn from the same parent
population. There is perhaps a hint that the Safronov numbers for
Class~I planets show a decreasing trend with metallicity in
Figure~\ref{fig:saffeh}, whereas the $\Theta$ values for Class~II
planets are independent of [Fe/H].

\subsection{Heavy element content versus stellar metallicity}
\label{sec:cores}

Transiting planet discoveries have challenged our theoretical
understanding of these objects from the very beginning. The inflated
radius of HD~209458b, argued to be due to an overlooked internal heat
source in the planet \citep[see, e.g.,][and references
therein]{Fabrycky:07}, still poses a problem for modelers, and this
first example is now joined by several other oversized planets
indicating it is not an exception (see \S\,\ref{sec:results}).  On the
other end of the scale, HD~149026b was the first transiting giant
planet found to have a radius significantly \emph{smaller} than
predicted by standard theories \citep{Sato:05}. The implication is
that it must have a substantial fraction of heavy elements of perhaps
$\sim$70~M$_{\earth}$ (2/3 of its total mass), which is often assumed
to be in a core.  More recently HAT-P-3b \citep{Torres:07a} has also
been found to be enhanced in heavy elements on the basis of its small
size for the measured planet mass. In this case metals make up
$\sim$1/3 of the total mass.

\cite{Guillot:06} and also \cite{Burrows:07} have found evidence that
the heavy element content correlates with the metallicity of the
parent star, a trend that was not anticipated by theory. These studies
were based, respectively, on the 9 and 14 transiting planets known at
the time. The larger sample now available warrants a second look at
this possible correlation, for its potential importance for our
understanding of planet formation.

Here we have estimated the heavy element content $M_{\rm Z}$ of each
planet using the recent models by \cite{Fortney:07}, which include the
effects of irradiation from the central star. The mass in heavy
elements was calculated from the measured mass of the planet, its
radius, the orbital semimajor axis, and the age of the system as
derived above from stellar evolution models. The uncertainty in these
estimates is often very large, due mostly to errors in the measured
values of $R_p$ and especially the age. In ten cases the models
indicate no metal enhancement at all, and some of these are in fact
``inflated'' hot Jupiters. In a few other cases only an upper limit
can be placed. For HAT-P-2b the \cite{Fortney:07} models as published
do not provide a large enough range of core masses.  The results for
$M_{\rm Z}$ are listed in Table~\ref{tab:planetary}.  A comparison
with values from \cite{Guillot:06} and \cite{Burrows:07} for the
subset of planets in common indicates significant differences in some
cases, either in $M_{\rm Z}$ or its error.

\begin{figure}
\vskip -0.35in
\epsscale{1.3}
{\hskip -0.2in\plotone{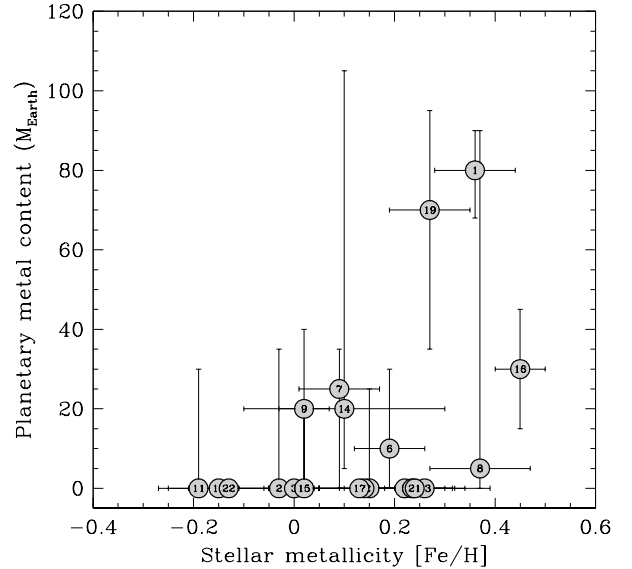}}
\vskip -0.3in

\figcaption[]{Heavy element content for transiting giant planets as a
function of the metallicity of the host star. HAT-P-2 is excluded (see
text).\label{fig:cores}}

\end{figure}

The mass in heavy elements is plotted against the stellar metallicity
in Figure~\ref{fig:cores}. The overall trend is similar to that
pointed out by previous authors, in the sense that the upper envelope
of the distribution appears to increase with [Fe/H]. It is natural to
expect that a higher stellar metallicity implies a higher-metallicity
protoplanetary disk. In the context of the core-accretion model of
planet formation, one would also naturally expect a more metal-rich
disk to lead to more metal-rich planets, and in this sense the
observed trend is in accordance with the core-accretion theory.

\section{Final remarks}
\label{sec:remarks}

Progress in understanding planet formation, structure, and evolution
depends to a large extent on an accurate knowledge of their physical
characteristics. This begins with an accurate knowledge of the
properties of the parent stars. The main contributions of this work
toward that goal are a re-analysis of all available transit light
curves with the same methodology and the uniform application of the
best possible observational constraints to infer the stellar mass and
radius. This includes, in particular, the constraint on the stellar
density through the light-curve parameter $a/R_{\star}$. We expect the
application of this technique to current and future ground-based or
space-based transit searches to be highly beneficial, especially when
parallax information for the candidates is unavailable.

Through the procedures described above we have obtained a more
homogeneous set of stellar and planetary parameters than previously
available, with error bars that are well understood and more
appropriate when searching for patterns and correlations among the
various quantities. The results have enabled us to explore the role of
metallicity in two of the more intriguing correlations between star
and planet properties: that of $M_p$ versus $P$, to which
\cite{Mazeh:05} first drew attention, and that of the Safronov number
versus equilibrium temperature \citep{Hansen:07}. We have also
re-examined the correlation between the stellar metallicity and the
heavy-element content of the planets.  We find a clear influence of
[Fe/H] in the $M_p$ versus $P$ diagram, with the planets in more
metal-poor systems being systematically more massive at a given
orbital period. This can be interpreted as evidence of a (mass and)
metallicity dependence of the migration process. Alternatively it may
be seen as indirect support for the correlation between core size and
[Fe/H] along with the idea that the presence of such cores slows down
or prevents complete evaporation of the planets in the extreme
radiation environments of these hot Jupiters. We find also that the
improved parameters for the present sample of transiting planets
support the recently proposed notion of at least two distinct classes
of hot Jupiters in terms of their Safronov numbers \citep{Hansen:07}.
Additionally, the average metallicity of Class~II planets appears to
be $\sim$0.2 dex higher than Class~I. Thus, chemical composition adds
another distinguishing characteristic to those proposed earlier.  A
further result of our work, which takes advantage of the larger sample
of transiting planets now available, is the confirmation of the
general trend of higher heavy-element content for planets with more
metal-rich host stars, originally advanced by \cite{Guillot:06} and
\cite{Burrows:07}.

We expect that the search for other significant correlations will be
made easier by the better accuracy and homogeneous character of the
properties of the present sample of transiting planets, and should
lead to a deeper understanding of their nature.

\acknowledgements

We thank Alex Sozzetti for providing information on TrES-3 and TrES-4
in advance of publication, and the referee for a helpful report. We
are grateful to Scott Gaudi for helpful conversations about the
analysis of transiting systems, and in particular for emphasizing the
importance of a uniform analysis. GT acknowledges partial support for
this work from NASA Origins grant NNG04LG89G. MJH acknowledges support
for this work from NASA Origins grant NNG06GH69G. This research has
made use of the VizieR service \citep{Ochsenbein:00} and of the SIMBAD
database, both operated at CDS, Strasbourg, France, as well as of
NASA's Astrophysics Data System Abstract Service, and of data products
from the Two Micron All Sky Survey, which is a joint project of the
University of Massachusetts and the Infrared Processing and Analysis
Center/California Institute of Technology, funded by NASA and the NSF.

\appendix
\section{Spectroscopic and photometric information from the literature}
\label{sec:app}

Below we describe in detail our compilation of the information
available in the literature for the atmospheric parameters of all
known transiting planets, as well as the sources for the light curves
we have used, and for the velocity semi-amplitudes that determine the
planet masses. For the atmospheric parameters ($T_{\rm eff}$, [Fe/H],
$\log g$) we consider only spectroscopic studies, and disregard
photometric determinations with few exceptions.  The values adopted,
which are listed in Table~\ref{tab:atmospheric}, are the result of a
careful combination of results for each planet rather than the
selection of a particular favorite study, and we have therefore deemed
it important to document our choices for future reference. For the
spectroscopic parameters the goal of this effort has been to arrive at
values that best represent the stellar properties based on current
knowledge. Combined with our application of uniform procedures for
deriving the stellar mass and radius based on state-of-the-art stellar
evolution models, we hope that the overall accuracy of the planet
parameters has been improved.

\noindent{\bf 1.~HD~149026:} Two consistent determinations of the
effective temperature are available for this star. One is from
\cite{Masana:06}, who apply a semi-empirical method based on a
spectral energy distribution fit (SEDF) calibrated against solar
analogs, and the other is from a spectral synthesis analysis by
\cite{Sato:05} carried out on high-resolution Subaru spectra using the
SME package \citep[Spectroscopy Made Easy,][]{Valenti:96, Valenti:05}.
We adopt the weighted average of the two, with a generous uncertainty
of 50~K. The iron abundance is adopted from \cite{Sato:05}, but with a
more conservative error of 0.08 dex. The surface gravity is taken from
the same source. The velocity semi-amplitude $K$ is adopted from
\cite{Butler:06}. We consider that study, based on 16 RV measurements
with the Keck telescope, to supersede the one by \cite{Sato:05}, which
used fewer Keck velocities (6) combined with 4 velocities with the
Subaru telescope, but obtained essentially the same result.
Additional Keck velocity measurements (many obtained during transit)
were reported recently by \cite{Wolf:07}, who also re-reduced the
earlier Keck observations of \cite{Butler:06} (for a total of
35). Their study focused on the Rossiter-McLaughlin effect, and did
not present a value for $K$. The photometric data used here are the
three St\"omgren $(b\!+\!y)/2$ light curves presented by
\cite{Sato:05}, the Sloan $g$ and $r$ light curves presented by
\cite{Charbonneau:06}, and the five $(b\!+\!y)/2$ light curves
presented by \cite{Winn:07e}.

\noindent{\bf 2.~HD~189733:} For $T_{\rm eff}$ we adopt the weighted
average of the results by \cite{Bouchy:05a} based on high-resolution
CORALIE spectra, and \cite{Masana:06}, with an uncertainty of 50~K.
Later papers by the first team quote values for $T_{\rm eff}$ (as well
as $\log g$ and [Fe/H]) said to be taken from \cite{Bouchy:05a}, but
which are not exactly the same for reasons that are unclear. We adopt
$\log g$ from the Bouchy work, as well as the metallicity, but with an
increased [Fe/H] error of 0.08 dex. A study by \cite{Gray:03} based on
lower resolution spectra (1.8~\AA) gave a somewhat lower temperature
(4939~K) along with a much lower value for [Fe/H] of $-0.37$; we do
not consider those results here. The only published determination of
$K$ is by \cite{Bouchy:05a}, which we adopt. Additional RV
measurements at Keck were obtained by \cite{Winn:06}, mostly during
transit for the purpose of studying the Rossiter-McLaughlin effect,
but no value of $K$ was reported. Ground-based light curves have been
obtained in a variety of passbands since the discovery (Str\"omgren
$(b\!+\!y)/2$, $V$, $I$, Sloan $z$).  High-quality observations with
HST in a passband with an effective wavelength similar to the $R$ band
were reported by \cite{Pont:07b}. These optical observations all
suffer to some degree from the fact that the star is relatively
active.  Spottedness causes irregularities in the photometry, which is
very useful to establish the rotational period of the star
\citep[see][]{Winn:07b, Henry:07}, but interferes with the accurate
determination of the light-curve parameters. By contrast,
\cite{Knutson:07} obtained photometry with the Spitzer Space Telescope
at the much longer wavelength of 8~$\mu$m, where spots have a
negligible effect.  Limb-darkening is virtually non-existent at this
wavelength as well, removing another complication. While the
ground-based, HST, and Spitzer light curve solutions all agree very
well within their uncertainties, for the reasons just described we
have chosen in this case to rely only on the Spitzer data, which we
have re-analyzed using the same methodology applied to all the other
transiting planets. A very similar (but less precise) radius ratio was
reported recently by \cite{Ehrenreich:07} based on additional Spitzer
observations at 3.6 and 5.8~$\mu$m.

\noindent{\bf 3.~HD~209458:} Ten independent determinations of $T_{\rm
eff}$ are available, which show excellent agreement: eight are from
spectroscopic studies \citep{Mazeh:00, Gonzalez:01, Mashonkina:01,
Sadakane:02, Heiter:03, Santos:04, Valenti:05}, and two others are
based on photometry use the IRFM \citep{Ramirez:04} and SEDF
\citep{Masana:06}.  We adopt the weighted average, with a conservative
uncertainty of 50~K. Similarly, we adopt the weighted average of the
nine [Fe/H] determinations from the spectroscopic sources, with an
error of 0.05 dex. The $\log g$ value we use here is also the weighted
average of the eight available measurements. Four values for the
semi-amplitude $K$ have been published. The early determinations by
\cite{Mazeh:00} and \cite{Henry:00} were superseded by more recent
studies based on many more RVs and refined analysis techniques by
\cite{Naef:04} (46 ELODIE and 141 CORALIE measurements over 5 years,
not all published) and \cite{Butler:06} (64 Keck measurements over 6
years). We adopt the weighted average of the results from the two new
sources. Our light curve fits are based on the HST/STIS photometry of
\cite{Brown:01}.  In this case, given the high signal-to-noise ratio
of the data, we fitted for the quadratic limb-darkening coefficients
along with all of the other parameters.

\noindent{\bf 4.~OGLE-TR-10:} A spectroscopic study by
\cite{Santos:06} used 4 co-added UVES spectra to derive $T_{\rm eff}$,
[Fe/H], and $\log g$. An earlier study by \cite{Bouchy:05b} used a
subset of these spectra and somewhat different analysis techniques,
and obtained a significantly hotter temperature and a higher
metallicity. They also reported an unusually large surface gravity for
the temperature they obtained ($\log g = 4.70$, for $T_{\rm eff} =
6220$~K). Because these results may be affected by the strong
correlations often present between all these quantities, we have
preferred not to use the \cite{Bouchy:05b} values here. Another recent
study by \cite{Holman:07a} supersedes previous analyses by
\cite{Konacki:03a} and \cite{Konacki:05} based on similar material,
which, nevertheless gave essentially the same results with only
slightly different errors. The Santos and Holman values for $T_{\rm
eff}$ ($6075 \pm 86$~K and $5800 \pm 100$~K, respectively) are
somewhat discrepant (by 275~K, or 2.1$\sigma$). We note that the
direction of the difference is consistent with the sign of the
difference in the iron abundances of these two studies (Santos giving
$+0.28 \pm 0.10$, and Holman $0.0 \pm 0.2$), given the typical
correlation between these two quantities.  Both teams have presented
evidence supporting their $T_{\rm eff}$ determinations (a good fit to
the H$\alpha$ line profile in the case of Santos et al., and agreement
with the H$\beta$ profile as well as a consistency check using
line-depth ratios, for Holman et al.).  Under these circumstances we
simply adopt intermediate values (arithmetic averages) for $T_{\rm
eff}$, [Fe/H], and $\log g$, with increased uncertainties of 130~K,
0.15 dex, and 0.10 dex, respectively. Two consistent determinations of
$K$ have been reported: one by \cite{Bouchy:05b} based on 14 RVs from
UVES/FLAMES, and the other by \cite{Konacki:05}. We adopt the latter
results, because they are based on a combination of 9 Keck
measurements with the 14 RVs from \cite{Bouchy:05b}. The light-curve
parameters are taken from the work of \cite{Pont:07a}, who combine the
best photometry available.

\noindent{\bf 5.~OGLE-TR-56:} The adopted temperature, metallicity,
and surface gravity for this star are the weighted averages of three
fairly consistent spectroscopic determinations \citep{Konacki:03b,
Bouchy:05b, Santos:06}. Conservative errors of 100~K, 0.10 dex, and
0.16 dex are assigned to those quantities, respectively. Two
determinations of the semi-amplitude $K$ are considered here: one by
\cite{Torres:04}, who used 11 Keck RVs and whose study supersedes
\cite{Konacki:03b}, and one by \cite{Bouchy:05b}, based on 8 UVES and
5 HARPS measurements. We adopt the weighted average of the two
results. Our light curve fits are based on the $V$ and $R$ data of
\cite{Pont:07a}.

\noindent{\bf 6.~OGLE-TR-111:} The atmospheric parameters for this
system are adopted from the spectroscopic study by \cite{Santos:06},
based on 6 UVES spectra, which supersedes those in the discovery paper
by \cite{Pont:04} that used only a subset of the same spectra. The
errors seem realistic, so we adopt those as well. The only
determination of $K$ available is from \cite{Pont:04}. Our photometric
solutions are based on the two $I$-band light curves of
\cite{Winn:07f}.

\noindent{\bf 7.~OGLE-TR-113:} Three spectroscopic determinations of
$T_{\rm eff}$, [Fe/H], and $\log g$ give very consistent results
\citep{Bouchy:04, Konacki:04, Santos:06}, and we adopt the weighted
average. The formal uncertainties are very small because of the
(likely accidental) good agreement; to be conservative we have
increased them to 75~K, 0.08 dex, and 0.22 dex, respectively. For $K$
we have taken the weighted average from the first two sources above,
which are based, respectively, on 8 spectra from UVES and 7 from
Keck. Our light curve fits use the $R$ band photometry of
\cite{Gillon:06}.

\noindent{\bf 8.~OGLE-TR-132:} The temperature, metallicity, and
surface gravity are taken from \cite{Gillon:07a}. This high-resolution
study, based on UVES spectra, supersedes an earlier one by
\cite{Bouchy:04}, which had very large uncertainties. The velocity
semi-amplitude is adopted from the work of \cite{Moutou:04}, who
obtained a $K$ value some 20\% larger than in the discovery paper of
\cite{Bouchy:04} despite having used the same 5 original RV
measurements. The difference is due solely to the improved ephemeris
provided by \cite{Moutou:04} based on a new, high-quality light curve
gathered with the VLT, compared to the original ephemeris from the
OGLE photometry. Our light curve fits use the VLT $R$-band photometry
of \cite{Gillon:07a}.

\noindent{\bf 9.~TrES-1:} Four high-resolution spectroscopic analyses
of this star giving consistent results have been published by
\cite{Alonso:04}, \cite{Sozzetti:04}, \cite{Laughlin:05}, and
\cite{Santos:06}. We adopt the weighted averages of those studies.
The metallicity from \cite{Alonso:04} was assigned here an uncertainty
(not originally reported) of 0.2 dex based on previous experience with
the same type of material. As before, we have chosen more conservative
errors of 50~K for $T_{\rm eff}$ and 0.05 dex for [Fe/H], to account
in the latter case for the fact that \cite{Santos:06} find a hint of
small systematic differences with \cite{Sozzetti:04}. The mean
metallicity is close to solar: [Fe/H] $= +0.02 \pm 0.05$.  A claim has
been made by \cite{Strassmeier:04} of a much lower metallicity for
TrES-1 of [Fe/H] $= -0.6$ (no error given). This is so discrepant
compared to other four determinations that it is most likely
incorrect. For $K$, the only value appearing in the literature is that
of \cite{Alonso:04}, based on 8 Keck spectra, which we adopt.
\cite{Laughlin:05} made 5 additional RV measurements, also with Keck,
and combined them with the Alonso velocities to improve the planet
mass.  However, no value for $K$ was reported and it is not possible
to recover it accurately from other published quantities. Our light
curve fits use the Sloan $z$-band photometry of \cite{Winn:07d}.

\noindent{\bf 10.~TrES-2:} A single source is available for the
atmospheric parameters \citep{Sozzetti:07}, and those values are
adopted here. Although the published uncertainty in $T_{\rm eff}$ is
only 50~K, a number of checks on the temperature are presented that
suggest the value is accurate, and thus we accept that error. The
semi-amplitude $K$ is adopted from the discovery paper by
\cite{ODonovan:06}. Our photometric solutions rely on the $z$-band
light curves of \cite{Holman:07b}.

\noindent{\bf 11.~TrES-3:} The atmospheric properties were taken from
Sozzetti et al.\ (in preparation), while the semi-amplitude $K$ and
the light-curve parameters are adopted from the discovery paper by
\cite{ODonovan:07}.

\noindent{\bf 12.~TrES-4:} The atmospheric properties were taken from
Sozzetti et al.\ (in preparation), while the semi-amplitude $K$ and
the light-curve parameters are adopted from the discovery paper by
\cite{Mandushev:07}.

\noindent{\bf 13.~WASP-1:} The spectroscopic investigation by
\cite{Stempels:07} based on NOT/FIES spectra supersedes the results in
the discovery paper by \cite{CollierCameron:07}, and in fact the new
study claims that the SOPHIE spectra in the old one were compromised
by scattered light and uncertain normalization. Despite this, the
results for the atmospheric quantities are quite similar.  We adopt
the parameters from the new study, although with more conservative
uncertainties of 75~K, 0.08 dex, and 0.16 dex for $T_{\rm eff}$,
[Fe/H], and $\log g$. The only RV analysis available is that by
\cite{CollierCameron:07}, based on 7 RV measurements with SOPHIE.  Our
light curve fits are based on the $z$-band photometry of
\cite{Charbonneau:07}.

\noindent{\bf 14.~WASP-2:} For this system there are no spectroscopic
studies beyond the one in the discovery paper by
\cite{CollierCameron:07}, in which the uncertainties for $T_{\rm eff}$
and $\log g$ are rather large. Nevertheless, the temperature appears
to be accurate as shown by a consistency check obtained using the
IRFM.  The authors state that the abundances are not substantially
different from solar, and they appear to adopt, according to the
caption of their Table~3, a value of [Fe/H] $= +0.1 \pm 0.2$. This may
simply be a representative abundance, but it is a reasonable value
since it is close to the average metallicity of the other transiting
planets. We accept it for lack of a better determination, along with
$T_{\rm eff}$ and $\log g$ as reported. We take also the $K$ value
from this paper, which is based on 9 RV measurements with SOPHIE.  Our
light curve fits use the $z$-band photometry of \cite{Charbonneau:07}.

\noindent{\bf 15.~XO-1:} The discovery paper by \cite{McCullough:06}
presents the only spectroscopic study of this system. The atmospheric
parameters are based on HJS spectra from the 2.7m telescope at
McDonald Observatory, analyzed with SME. We adopt the temperature and
metallicity from this source, with increased uncertainties of 75~K and
0.08 dex. The $\log g$ value is adopted also as given there. The
semi-amplitude $K$ derived by these authors is based on 10 RV
measurements from HET and the 2.7m (HJS). Our light curve fits are
based on the $z$-band photometry of \cite{Holman:06}.

\noindent{\bf 16.~XO-2:} The values of $T_{\rm eff}$, [Fe/H], and
$\log g$ derived in the discovery paper by \cite{Burke:07} are adopted
here as published, and are based on two spectra collected with the HET
for this northern component of a 31\arcsec\ visual binary. The
logarithmic iron abundance, $+0.45 \pm 0.02$, is found to be very
similar to that obtained for the visual companion, which is $+0.47 \pm
0.02$. We adopt conservative uncertainties of 80~K for $T_{\rm eff}$
and 0.05 dex for [Fe/H]. The $K$ value from this study is based on 10
RV measurements, and is adopted for our work. The light curve
parameters adopted are also taken from the discovery paper.

\noindent{\bf 17.~HAT-P-1:} A single determination of the atmospheric
parameters has been published, in the discovery paper by
\cite{Bakos:07a}. It is based on an SME analysis of a Keck spectrum.
The star is the easterly and fainter component of an 11\arcsec\ visual
binary known as ADS~16402. \cite{Bakos:07a} also analyzed the other
component spectroscopically, and used both sets of parameters for this
physically associated and coeval pair to better constrain the mass and
radius using evolutionary models. In this process they noted some
inconsistencies between the models and the observations (a steeper
slope in the H-R diagram than predicted by theory), which advise
caution in using the atmospheric parameters. We adopt the values as
published, but with increased uncertainties of 120~K, 0.08 dex, and
0.15 dex for $T_{\rm eff}$, [Fe/H], and $\log g$, respectively. The
velocity semi-amplitude is also taken from \cite{Bakos:07a}, and is
based on 9 velocities from Keck and 4 from Subaru. Our light curve
solutions are based on the $z$-band photometry of \cite{Winn:07c}.

\noindent{\bf 18.~HAT-P-2:} For convenience we use the HAT designation
here, although the star is also referred to as HD~147506. We adopt the
atmospheric parameters from the discovery paper by \cite{Bakos:07b},
the only one presenting a detailed spectroscopic analysis. An
independent study by \cite{Loeillet:07} based on SOPHIE spectra did
not yield better estimates, according to the authors, although it did
confirm the [Fe/H] value. The velocity semi-amplitude in the discovery
paper has been superseded by the determination by \cite{Loeillet:07},
who included the original velocities from \cite{Bakos:07b} along with
63 new measurements obtained with SOPHIE.  The latter were intended to
study the Rossiter-McLaughlin effect, and were therefore taken mostly
during transit. The authors modeled the complete velocity curve, and
improved not only $K$ but also the eccentricity $e$ and longitude of
periastron $\omega$. We adopt those elements in this work. In another
study by \cite{Winn:07a} a total of 97 new spectra from Keck were
obtained, also to investigate the Rossiter-McLaughlin phenomenon. The
authors focussed mainly on that effect and did not report a new value
of $K$. The light curve parameters reported here are based on the
$z$-band photometry of \cite{Bakos:07b}.

\noindent{\bf 19.~HAT-P-3:} The atmospheric parameters are adopted
from the only spectroscopic analysis so far, which is the discovery
paper by \cite{Torres:07a} that is based on an SME analysis of a Keck
spectrum.  The uncertainties adopted for $T_{\rm eff}$, [Fe/H], and
$\log g$, which are 80~K, 0.08 dex, and 0.08 dex, are more
conservative than the formal errors returned by SME, as described
there. The value of $K$ is taken from the same source, as are the
light curve parameters.

\noindent{\bf 20.~HAT-P-4:} As in the previous case, the atmospheric
parameters, the $K$ value, and the light-curve parameters are adopted
from the discovery paper \citep{Kovacs:07}. The uncertainties in
$T_{\rm eff}$ and [Fe/H] are the same as for HAT-P-3.

\noindent{\bf 21.~HAT-P-5:} All atmospheric, spectroscopic, and
light-curve parameters are taken from the discovery paper
\citep{Bakos:07c}.

\noindent{\bf 22.~HAT-P-6:} All parameters are taken from the
discovery paper \citep{Noyes:07}.

\noindent{\bf 23.~GJ~436:} Only one valiant attempt has been made to
determine the atmospheric parameters of this M dwarf spectroscopically
\citep{Maness:07}. It is based on both low- and high-resolution
observations and a comparison with synthetic spectra. The difficulties
of this kind of measurement are evidenced by the disagreements the
authors obtain from their low- and high-resolution material, in both
effective temperature (3500~K and 3200~K, respectively) and surface
gravity (5.0 and 4.0 dex). They attribute this to shortcomings in our
knowledge of the molecular opacities, mainly of TiO. They adopt a
compromise value of $T_{\rm eff} = 3350 \pm 300$~K for a fixed surface
gravity of $\log g = 5.0$, and a metallicity set to the solar
value. The latter assumption is supported by a photometric
determination of [Fe/H] $= -0.03 \pm 0.20$ by \cite{Bonfils:05}, based
on the absolute $K$-band magnitude and the $V\!-\!K$ color. These are
the values we adopt here. The velocity semi-amplitude by
\cite{Maness:07} uses 59 measurements from Keck, and supersedes the
determination in the discovery paper by \cite{Butler:04}. Their
orbital model includes a linear drift in the velocities presumably
caused by a more distant orbiting companion.  \cite{Demory:07} use the
same velocities, also solving for a residual linear trend, and they
additionally incorporate an accurate timing measurement of the transit
and another of the secondary eclipse based on Spitzer
observations. They report a $K$ semi-amplitude similar to that of
\cite{Maness:07}, but give no uncertainty. We adopt the
\cite{Maness:07} value here. For the light curve parameters we adopt a
weighted average of the values reported by (or reconstructed from) the
ground-based or Spitzer-based studies of \cite{Gillon:07b},
\cite{Deming:07}, and \cite{Gillon:07c}: $a/R_{\star} = 13.34 \pm
0.58$, $R_p/R_{\star} = 0.0834 \pm 0.0007$, and $b = 0.848 \pm 0.010$.

\clearpage

\begin{deluxetable*}{clccccc}
\tabletypesize{\scriptsize}
\tablewidth{0pc}
%\rotate
\tablecaption{Adopted atmospheric properties and orbital semi-amplitudes of the host stars.\label{tab:atmospheric}}
\tablehead{\colhead {} & \colhead{} & \colhead{$V$} & \colhead{$T_{\rm eff}$} &
\colhead{[Fe/H]} & \colhead{$\log g_{\rm spec}$\tablenotemark{a}} & \colhead{$K$} \\
\colhead{\#} & \colhead{~~~~~~~~~Name~~~~~~~~~} & \colhead{(mag)} &
\colhead{(K)} & \colhead{(dex)} & \colhead{(cgs)} & \colhead{(\ms)}}
\startdata
\phn1 & HD 149026\dotfill   & \phn$8.16~\pm~0.01$ &    6160~$\pm$~\phn50  &  $+$0.36~$\pm$~0.08   &  4.26~$\pm$~0.07  &   43.2~$\pm$~2.3\phn     \\
\phn2 & HD 189733\tablenotemark{b}\dotfill   & \phn$7.67~\pm~0.03$ &    5040~$\pm$~\phn50  &  $-$0.03~$\pm$~0.08   &  4.53~$\pm$~0.14  &  205~$\pm$~6\phn\phn     \\
\phn3 & HD 209458\dotfill   & \phn$7.65~\pm~0.01$ &    6065~$\pm$~\phn50  &  \phs0.00~$\pm$~0.05  &  4.42~$\pm$~0.04  &  84.67~$\pm$~0.70\phn    \\
\phn4 & OGLE-TR-10\dotfill  & \phm{$i$~}14.93~$I$    &    5950~$\pm$~130     &  $+$0.15~$\pm$~0.15   &  4.50~$\pm$~0.10  &  80~$\pm$~17             \\
\phn5 & OGLE-TR-56\dotfill  & \phm{$i$~}15.30~$I$    &    6050~$\pm$~100     &  $+$0.22~$\pm$~0.10   &  4.22~$\pm$~0.16  &  225~$\pm$~27\phn        \\
\phn6 & OGLE-TR-111\dotfill & \phm{$i$~}15.55~$I$    &    5040~$\pm$~\phn80  &  $+$0.19~$\pm$~0.07   &  4.51~$\pm$~0.36  &  78~$\pm$~14             \\
\phn7 & OGLE-TR-113\dotfill & \phm{$i$~}14.42~$I$    &    4790~$\pm$~\phn75  &  $+$0.09~$\pm$~0.08   &  4.51~$\pm$~0.22  &  267~$\pm$~34\phn        \\
\phn8 & OGLE-TR-132\dotfill & \phm{$i$~}15.72~$I$    &    6210~$\pm$~\phn75  &  $+$0.37~$\pm$~0.10   &  4.51~$\pm$~0.27  &  167~$\pm$~18\phn        \\
\phn9 & TrES-1\dotfill      & $11.76~\pm~0.01$    &    5230~$\pm$~\phn50  &  $+$0.02~$\pm$~0.05   &  4.50~$\pm$~0.04  &  115.2~$\pm$~6.2\phn\phn \\
10 & TrES-2\dotfill         & $11.41~\pm~0.01$    &    5850~$\pm$~\phn50  &  $-$0.15~$\pm$~0.10   &  4.4~$\pm$~0.1    &  181.3~$\pm$~2.6\phn\phn \\
11 & TrES-3\dotfill         & $12.40~\pm~0.01$    &    5650~$\pm$~\phn75  &  $-$0.19~$\pm$~0.08   &  4.4~$\pm$~0.1    &  378.4~$\pm$~9.9\phn\phn \\
12 & TrES-4\dotfill         & $11.59~\pm~0.01$    &    6200~$\pm$~\phn75  &  $+$0.14~$\pm$~0.09   &  4.0~$\pm$~0.1    &  97.4~$\pm$~7.2\phn      \\
13 & WASP-1\dotfill         & $11.68~\pm~0.05$    &    6110~$\pm$~\phn75  &  $+$0.26~$\pm$~0.08   &  4.28~$\pm$~0.16  &  115~$\pm$~11\phn        \\
14 & WASP-2\dotfill         & $11.88~\pm~0.10$    &    5200~$\pm$~200     &  $+$0.1~$\pm$~0.2     &  4.3~$\pm$~0.3    &  155~$\pm$~7\phn\phn     \\
15 & XO-1\dotfill           & $11.19~\pm~0.03$    &    5750~$\pm$~\phn75  &  $+$0.02~$\pm$~0.08   &  4.53~$\pm$~0.06  &  116~$\pm$~9\phn\phn     \\
16 & XO-2\dotfill           & $11.18~\pm~0.03$    &    5340~$\pm$~\phn80  &  $+$0.45~$\pm$~0.05   &  4.48~$\pm$~0.05  &  85~$\pm$~8\phn          \\
17 & HAT-P-1\dotfill        & $10.33~\pm~0.05$    &    5975~$\pm$~120     &  $+$0.13~$\pm$~0.08   &  4.45~$\pm$~0.15  &  60.3~$\pm$~2.1\phn      \\
18 & HAT-P-2\tablenotemark{c}\dotfill        & \phn$8.71~\pm~0.01$ &    6290~$\pm$~110     &  $+$0.12~$\pm$~0.08   &  4.22~$\pm$~0.14  &  968.6~$\pm$~8.3\phn\phn \\
19 & HAT-P-3\dotfill        & $11.55~\pm~0.05$    &    5185~$\pm$~\phn80  &  $+$0.27~$\pm$~0.08   &  4.61~$\pm$~0.08  &  89.1~$\pm$~2.0\phn      \\
20 & HAT-P-4\dotfill        & $11.22~\pm~0.03$    &    5860~$\pm$~\phn80  &  $+$0.24~$\pm$~0.08   &  4.36~$\pm$~0.11  &  81.1~$\pm$~1.9\phn      \\
21 & HAT-P-5\dotfill        & $12.00~\pm~0.06$    &    5960~$\pm$~100     &  $+$0.24~$\pm$~0.15   &  4.0~$\pm$~0.2    &   138~$\pm$~14\phn      \\
22 & HAT-P-6\dotfill        & $10.44~\pm~0.04$    &    6570~$\pm$~\phn80  &  $-$0.13~$\pm$~0.08   &  3.84~$\pm$~0.12  & 115.5~$\pm$~4.2\phn\phn  \\
23 & GJ 436\tablenotemark{d}\dotfill         & $10.67~\pm~0.01$    &    3350~$\pm$~300     &  $-$0.03~$\pm$~0.20   &  4.5~$\pm$~0.5    &  18.34~$\pm$~0.52\phn \\ [-2.0ex]
\enddata

\tablenotetext{a}{These spectroscopically determined surface gravities
are generally less reliable than those derived from our stellar
evolution modeling (see Table~\ref{tab:stellar}), and are included
here only for completeness.}
\tablenotetext{b}{The large photometric uncertainty in $V$ for such a
bright star is due to the variability of the object \citep[see,
e.g.,][]{Winn:07b}.}
\tablenotetext{c}{The orbit of the planet is not circular, and has $e
= 0.5170_{-0.0010}^{+0.0017}$ and $\omega = 189.1_{-0.3}^{+0.4}$
degrees \citep{Loeillet:07}.}
\tablenotetext{d}{The orbit of the planet is not circular, and has $e
= 0.14 \pm 0.01$ and $\omega = 350$ degrees \citep{Demory:07}. No
uncertainty was reported for $\omega$. The orbital fit includes a
linear trend presumably due to an outer orbiting companion.}
\tablecomments{See Appendix for the sources of these
determinations. Apparent magnitudes for the OGLE stars are in the $I$
band rather than $V$, and have estimated uncertainties of 0.05 mag.}
\end{deluxetable*}

\clearpage

\begin{deluxetable*}{cllcccccc}
\tabletypesize{\scriptsize}
\tablewidth{0pc}
%\rotate
\tablecaption{Light curve parameters for transiting planet systems.\label{tab:lcfits}}
\tablehead{\colhead{} & \colhead{} & \colhead{Period} & \colhead{} & \colhead{} &
\colhead{} & \colhead{$i$} & \colhead{$\rho_{\star}$\tablenotemark{a}} & \colhead{Depth in $V$} \\
\colhead{\#} & \colhead{~~~~~~~~~Name~~~~~~~~~} & \colhead{(days)} & \colhead{$R_p/R_{\star}$} &
\colhead{$a/R_{\star}$} & \colhead{$b \equiv a\cos i/R_{\star}$} & \colhead{(deg)} & \colhead{(g~cm$^{-3}$)} & \colhead{(mmag)}}
\startdata
\phn1 & HD 149026\dotfill   & 2.87598    & $0.0491^{+0.0018}_{-0.0005}$     & $7.11^{+0.03}_{-0.81}$  & $0.00^{+0.33}_{-0.00}$       & $90.0^{+0.0}_{-3.0}$    &  $0.822_{-0.25}^{+0.011}$ & \phn3.0  \\ [+1.5ex]
\phn2 & HD 189733\dotfill   & 2.218573   & $0.15463^{+0.00022}_{-0.00022}$  & $8.81^{+0.06}_{-0.06}$  & $0.680^{+0.005}_{-0.005}$    & $85.58^{+0.06}_{-0.06}$ &  $2.629_{-0.053}^{+0.054}$ & 25.3  \\ [+1.5ex]
\phn3 & HD 209458\dotfill   & 3.524746   & $0.12086^{+0.00010}_{-0.00010}$  & $8.76^{+0.04}_{-0.04}$  & $0.507^{+0.005}_{-0.005}$    & $86.71^{+0.05}_{-0.05}$ &  $1.024_{-0.014}^{+0.014}$ & 17.0  \\ [+1.5ex]
\phn4 & OGLE-TR-10\dotfill  & 3.101278   & $0.110^{+0.002}_{-0.002}$        & $8.07^{+0.44}_{-0.69}$  & $0.00^{+0.54}_{-0.00}$       & $90.0^{+0.0}_{-3.9}$    &  $1.03_{-0.24}^{+0.18}$ & 15.2  \\ [+1.5ex]
\phn5 & OGLE-TR-56\dotfill  & 1.2119189  & $0.1027^{+0.0019}_{-0.0019}$     & $3.74^{+0.19}_{-0.15}$  & $0.817^{+0.016}_{-0.031}$    & $77.60^{+0.91}_{-1.0}$  &  $0.674_{-0.078}^{+0.11}$ & 10.0  \\ [+1.5ex]
\phn6 & OGLE-TR-111\dotfill & 4.01610    & $0.1299^{+0.0010}_{-0.0013}$     & $12.09^{+0.45}_{-0.45}$ & $0.360^{+0.048}_{-0.17}$     & $88.25^{+0.83}_{-0.30}$ &  $2.07_{-0.22}^{+0.24}$ & 20.4  \\ [+1.5ex]
\phn7 & OGLE-TR-113\dotfill & 1.4324752  & $0.1450^{+0.0016}_{-0.0005}$     & $6.38^{+0.03}_{-0.27}$  & $0.24^{+0.06}_{-0.18}$       & $87.80^{+1.6}_{-0.62}$  &  $2.395_{-0.29}^{+0.034}$ & 26.0  \\ [+1.5ex]
\phn8 & OGLE-TR-132\dotfill & 1.689868   & $0.0932^{+0.0015}_{-0.0020}$     & $4.76^{+0.57}_{-0.30}$  & $0.560^{+0.046}_{-0.22}$     & $83.4^{+2.9}_{-1.3}$    &  $0.72_{-0.13}^{+0.29}$ & 10.0  \\ [+1.5ex]
\phn9 & TrES-1\dotfill      & 3.030065   & $0.13578^{+0.00035}_{-0.00032}$  & $10.52^{+0.02}_{-0.18}$ & $0.00^{+0.19}_{-0.00}$       & $90.0^{+0.0}_{-1.1}$    &  $2.400_{-0.12}^{+0.014}$ & 23.2  \\ [+1.5ex]
10 & TrES-2\dotfill         & 2.47063    & $0.1253^{+0.0010}_{-0.0010}$     & $7.62^{+0.11}_{-0.11}$  & $0.8540^{+0.0062}_{-0.0062}$ & $83.57^{+0.14}_{-0.14}$ &  $1.372_{-0.059}^{+0.061}$ & 13.6  \\ [+1.5ex]
11 & TrES-3\dotfill         & 1.30619    & $0.1660^{+0.0024}_{-0.0024}$     & $6.06^{+0.10}_{-0.10}$  & $0.8277^{+0.0097}_{-0.0097}$ & $82.15^{+0.21}_{-0.21}$ &  $2.47_{-0.12}^{+0.12}$ & 24.6  \\ [+1.5ex]
12 & TrES-4\dotfill         & 3.553945   & $0.09903^{+0.00088}_{-0.00088}$  & $6.03^{+0.13}_{-0.13}$  & $0.755^{+0.015}_{-0.015}$    & $82.81^{+0.33}_{-0.33}$ &  $0.328_{-0.021}^{+0.022}$ & \phn9.8  \\ [+1.5ex]
13 & WASP-1\dotfill         & 2.519961   & $0.1025^{+0.0007}_{-0.0007}$     & $5.69^{+0.03}_{-0.21}$  & $0.00^{+0.27}_{-0.00}$       & $90.0^{+0.0}_{-2.9}$    &  $0.549_{-0.059}^{+0.009}$ & 13.2  \\ [+1.5ex]
14 & WASP-2\dotfill         & 2.152226   & $0.1310^{+0.0013}_{-0.0013}$     & $7.95^{+0.32}_{-0.20}$  & $0.724^{+0.017}_{-0.028}$    & $84.81^{+0.35}_{-0.27}$ &  $2.05_{-0.15}^{+0.26}$ & 17.7  \\ [+1.5ex]
15 & XO-1\dotfill           & 3.941534   & $0.1326^{+0.0004}_{-0.0005}$     & $11.55^{+0.03}_{-0.45}$ & $0.240^{+0.045}_{-0.14}$     & $88.81^{+0.70}_{-0.30}$ &  $1.877_{-0.21}^{+0.015}$ & 21.8  \\ [+1.5ex]
16 & XO-2\dotfill           & 2.615857   & $0.10395^{+0.00090}_{-0.00085}$  & $8.2^{+0.1}_{-0.2}$     & $0.158^{+0.11}_{-0.085}$     & $88.90^{+0.60}_{-0.75}$ &  $1.525_{-0.11}^{+0.057}$ & 13.5  \\ [+1.5ex]
17 & HAT-P-1\dotfill        & 4.46543    & $0.1124^{+0.0007}_{-0.0007}$     & $10.47^{+0.23}_{-0.23}$ & $0.712^{+0.017}_{-0.017}$    & $86.11^{+0.18}_{-0.18}$ &  $1.089_{-0.070}^{+0.074}$ & 13.1  \\ [+1.5ex]
18 & HAT-P-2\dotfill        & 5.633320   & $0.06840^{+0.00087}_{-0.00073}$  & $10.12^{+0.03}_{-0.99}$ & $0.00^{+0.44}_{-0.00}$       & $90.0^{+0.0}_{-3.4}$    &  $0.618_{-0.16}^{+0.006}$ & \phn5.9  \\ [+1.5ex]
19 & HAT-P-3\dotfill        & 2.899703   & $0.1109^{+0.0025}_{-0.0022}$     & $10.59^{+0.66}_{-0.84}$ & $0.51^{+0.11}_{-0.13}$       & $87.24^{+0.69}_{-0.69}$ &  $2.67_{-0.59}^{+0.54}$ & 14.3  \\ [+1.5ex]
20 & HAT-P-4\dotfill        & 3.056536   & $0.08200^{+0.00044}_{-0.00044}$  & $6.04^{+0.03}_{-0.18}$  & $0.01^{+0.23}_{-0.01}$       & $89.91^{+0.09}_{-2.2}$  &  $0.446_{-0.039}^{+0.007}$ & \phn8.5  \\ [+1.5ex]
21 & HAT-P-5\dotfill        & 2.788491   & $0.1106^{+0.0006}_{-0.0006}$     & $7.50^{+0.19}_{-0.19}$  & $0.425^{+0.045}_{-0.050}$    & $86.75^{+0.44}_{-0.44}$ &  $1.027_{-0.076}^{+0.081}$ & 14.6  \\ [+1.5ex]
22 & HAT-P-6\dotfill        & 3.852985   & $0.09338^{+0.00053}_{-0.00053}$  & $7.69^{+0.22}_{-0.22}$  & $0.602^{+0.030}_{-0.030}$    & $85.51^{+0.35}_{-0.35}$ &  $0.580_{-0.048}^{+0.052}$ & \phn9.7  \\ [+1.5ex]
23 & GJ 436\dotfill         & 2.64385    & $0.0834^{+0.0007}_{-0.0007}$     & $13.34^{+0.58}_{-0.58}$ & $0.848^{+0.010}_{-0.010}$    & $86.36^{+0.16}_{-0.17}$ &  $6.43_{-0.80}^{+0.87}$ & \phn6.5  \\ [-1.0ex]
\enddata

\tablenotetext{a}{Stellar mean density computed directly from
$a/R_{\star}$ and $P$ using $\rho_{\star} = {3\pi\over G
P^2}(a/R_{\star})^3$ \citep[see][]{Winn:07e}, ignoring the very
small corrective term due to the planet.}

\end{deluxetable*}

\clearpage

\begin{deluxetable*}{clcccccc}
\tabletypesize{\scriptsize}
\tablewidth{0pc}
%\rotate
\tablecaption{Inferred properties of the host stars from stellar evolution models.\label{tab:stellar}}
\tablehead{\colhead {} & \colhead{} & \colhead{$M_{\star}$} & \colhead{$R_{\star}$} &
\colhead{$\log g_{\star}$} & \colhead{$L_{\star}$} & \colhead{$M_V$}& \colhead{Age} \\
\colhead{\#} & \colhead{~~~~~~~~~Name~~~~~~~~~} & \colhead{(M$_{\sun}$)} &
\colhead{(R$_{\sun}$)} & \colhead{(cgs)} & \colhead{(L$_{\sun}$)} & \colhead{(mag)}
& \colhead{(Gyr)}}
\startdata
\phn1 & HD 149026\dotfill   & $1.294_{-0.050}^{+0.060}$ & $1.368_{-0.083}^{+0.12}$  & $4.278_{-0.063}^{+0.045}$ & $2.43_{-0.35}^{+0.53}$    & $3.80_{-0.24}^{+0.18}$    & $ 1.9_{-0.9}^{+0.9}$ \\ [+1.5ex]
\phn2 & HD 189733\dotfill   & $0.806_{-0.048}^{+0.048}$ & $0.756_{-0.018}^{+0.018}$ & $4.587_{-0.015}^{+0.014}$ & $0.331_{-0.028}^{+0.029}$ & $6.20_{-0.10}^{+0.11}$    & $ 6.8_{-4.4}^{+5.2}$ \\ [+1.5ex]
\phn3 & HD 209458\dotfill   & $1.119_{-0.033}^{+0.033}$ & $1.155_{-0.016}^{+0.014}$ & $4.361_{-0.008}^{+0.007}$ & $1.622_{-0.10}^{+0.097}$  & $4.273_{-0.069}^{+0.078}$ & $ 3.1_{-0.7}^{+0.8}$ \\ [+1.5ex]
\phn4 & OGLE-TR-10\dotfill  & $1.14_{-0.12}^{+0.10}$    & $1.17_{-0.11}^{+0.13}$    & $4.358_{-0.082}^{+0.064}$ & $1.54_{-0.38}^{+0.52}$    & $4.34_{-0.35}^{+0.33}$    & $ 3.2_{-3.1}^{+4.0}$ \\ [+1.5ex]
\phn5 & OGLE-TR-56\dotfill  & $1.228_{-0.078}^{+0.072}$ & $1.363_{-0.086}^{+0.089}$ & $4.258_{-0.043}^{+0.043}$ & $2.24_{-0.40}^{+0.48}$    & $3.92_{-0.23}^{+0.23}$    & $ 3.2_{-1.3}^{+1.0}$ \\ [+1.5ex]
\phn6 & OGLE-TR-111\dotfill & $0.852_{-0.052}^{+0.058}$ & $0.831_{-0.040}^{+0.045}$ & $4.529_{-0.042}^{+0.038}$ & $0.401_{-0.061}^{+0.072}$ & $5.99_{-0.20}^{+0.20}$    & $ 8.8_{-6.6}^{+5.2}$ \\ [+1.5ex]
\phn7 & OGLE-TR-113\dotfill & $0.779_{-0.015}^{+0.017}$ & $0.774_{-0.011}^{+0.020}$ & $4.552_{-0.017}^{+0.009}$ & $0.296_{-0.018}^{+0.021}$ & $6.400_{-0.085}^{+0.099}$ & $13.2_{-2.4}^{+0.8}$\phn \\ [+1.5ex]
\phn8 & OGLE-TR-132\dotfill & $1.305_{-0.067}^{+0.075}$ & $1.32_{-0.12}^{+0.17}$    & $4.313_{-0.090}^{+0.063}$ & $2.35_{-0.49}^{+0.76}$    & $3.84_{-0.33}^{+0.26}$    & $ 1.2_{-1.1}^{+1.5}$ \\ [+1.5ex]
\phn9 & TrES-1\dotfill      & $0.878_{-0.040}^{+0.038}$ & $0.807_{-0.016}^{+0.017}$ & $4.567_{-0.015}^{+0.012}$ & $0.438_{-0.033}^{+0.035}$ & $5.847_{-0.095}^{+0.097}$ & $ 3.7_{-2.8}^{+3.4}$ \\ [+1.5ex]
10 & TrES-2\dotfill         & $0.983_{-0.063}^{+0.059}$ & $1.003_{-0.033}^{+0.033}$ & $4.427_{-0.021}^{+0.019}$ & $1.06_{-0.10}^{+0.10}$    & $4.77_{-0.11}^{+0.12}$    & $ 5.0_{-2.1}^{+2.7}$ \\ [+1.5ex]
11 & TrES-3\dotfill         & $0.915_{-0.031}^{+0.021}$ & $0.812_{-0.025}^{+0.014}$ & $4.581_{-0.012}^{+0.017}$ & $0.592_{-0.047}^{+0.065}$ & $5.46_{-0.13}^{+0.11}$    & $ 0.6_{-0.4}^{+2.0}$ \\ [+1.5ex]
12 & TrES-4\dotfill         & $1.394_{-0.056}^{+0.060}$ & $1.816_{-0.062}^{+0.065}$ & $4.064_{-0.021}^{+0.021}$ & $4.39_{-0.48}^{+0.53}$    & $3.16_{-0.14}^{+0.13}$    & $ 2.9_{-0.4}^{+0.4}$ \\ [+1.5ex]
13 & WASP-1\dotfill         & $1.301_{-0.047}^{+0.049}$ & $1.517_{-0.045}^{+0.052}$ & $4.190_{-0.022}^{+0.020}$ & $2.88_{-0.30}^{+0.36}$    & $3.63_{-0.14}^{+0.13}$    & $ 3.0_{-0.6}^{+0.6}$ \\ [+1.5ex]
14 & WASP-2\dotfill         & $0.89_{-0.12}^{+0.12}$    & $0.840_{-0.065}^{+0.062}$ & $4.537_{-0.046}^{+0.035}$ & $0.47_{-0.13}^{+0.16}$    & $5.78_{-0.36}^{+0.40}$    & $ 5.6_{-5.6}^{+8.4}$ \\ [+1.5ex]
15 & XO-1\dotfill           & $1.027_{-0.061}^{+0.057}$ & $0.934_{-0.032}^{+0.037}$ & $4.509_{-0.027}^{+0.018}$ & $0.86_{-0.10}^{+0.12}$    & $5.02_{-0.16}^{+0.14}$    & $ 1.0_{-0.9}^{+3.1}$ \\ [+1.5ex]
16 & XO-2\dotfill           & $0.974_{-0.034}^{+0.032}$ & $0.971_{-0.026}^{+0.027}$ & $4.452_{-0.022}^{+0.020}$ & $0.689_{-0.074}^{+0.082}$ & $5.33_{-0.14}^{+0.14}$    & $ 5.8_{-2.3}^{+2.8}$ \\ [+1.5ex]
17 & HAT-P-1\dotfill        & $1.133_{-0.079}^{+0.075}$ & $1.135_{-0.048}^{+0.048}$ & $4.382_{-0.030}^{+0.027}$ & $1.48_{-0.25}^{+0.25}$    & $4.38_{-0.19}^{+0.21}$    & $ 2.7_{-2.0}^{+2.5}$ \\ [+1.5ex]
18 & HAT-P-2\dotfill        & $1.308_{-0.078}^{+0.088}$ & $1.506_{-0.096}^{+0.13}$  & $4.199_{-0.053}^{+0.043}$ & $3.20_{-0.58}^{+0.82}$    & $3.50_{-0.27}^{+0.24}$    & $ 2.6_{-0.8}^{+0.8}$ \\ [+1.5ex]
19 & HAT-P-3\dotfill        & $0.928_{-0.054}^{+0.044}$ & $0.833_{-0.044}^{+0.034}$ & $4.564_{-0.032}^{+0.032}$ & $0.449_{-0.064}^{+0.071}$ & $5.84_{-0.19}^{+0.21}$    & $ 1.5_{-1.4}^{+5.4}$ \\ [+1.5ex]
20 & HAT-P-4\dotfill        & $1.248_{-0.12}^{+0.070}$  & $1.596_{-0.075}^{+0.060}$ & $4.127_{-0.027}^{+0.019}$ & $2.70_{-0.36}^{+0.37}$    & $3.74_{-0.16}^{+0.17}$    & $ 4.6_{-1.0}^{+2.2}$ \\ [+1.5ex]
21 & HAT-P-5\dotfill        & $1.157_{-0.081}^{+0.043}$ & $1.165_{-0.052}^{+0.046}$ & $4.368_{-0.031}^{+0.025}$ & $1.54_{-0.22}^{+0.24}$    & $4.33_{-0.17}^{+0.19}$    & $ 2.6_{-1.4}^{+2.1}$ \\ [+1.5ex]
22 & HAT-P-6\dotfill        & $1.290_{-0.066}^{+0.064}$ & $1.463_{-0.063}^{+0.069}$ & $4.218_{-0.030}^{+0.027}$ & $3.59_{-0.46}^{+0.50}$    & $3.36_{-0.16}^{+0.16}$    & $ 2.3_{-0.6}^{+0.5}$ \\ [+1.5ex]
23 & GJ 436\tablenotemark{a}\dotfill         & $0.452_{-0.012}^{+0.014}$ & $0.464_{-0.011}^{+0.009}$ & $4.843_{-0.011}^{+0.018}$ & $0.0260_{-0.0017}^{+0.0014}$ & $10.244_{-0.082}^{+0.082}$ & $6.0_{-5.0}^{+4.0}$ \\ [-1.0ex]
\enddata

\tablenotetext{a}{The value of $M_V$ is biased due to missing opacity
sources in the models (see text), and the nominal age is only
indicative since it is essentially unconstrained for this unevolved M
dwarf.}

\end{deluxetable*}

\clearpage

\begin{deluxetable*}{clccccc}
\tabletypesize{\scriptsize}
\tablewidth{0pc}
%\rotate
\tablecaption{Angular diameters and parallaxes for transiting planet hosts.\label{tab:phi}}

\tablehead{\colhead {} & \colhead{} & \colhead{$\phi_{\rm mod}$} & \colhead{$\phi_{\rm SB}$} &
\colhead{Distance} & \colhead{$\pi_{\rm mod}$} & \colhead{$\pi_{\rm HIP}$} \\
\colhead{\#} & \colhead{~~~~~~~~~Name~~~~~~~~~} & \colhead{(mas)} &
\colhead{(mas)} & \colhead{(pc)} & \colhead{(mas)} & \colhead{(mas)}
}
\startdata
\phn1 & HD 149026\dotfill   &  0.171~$\pm$~0.021   &  0.1747~$\pm$~0.0021 & 74.4~$\pm$~7.2\phn       &  13.43~$\pm$~1.29\phn & 12.68~$\pm$~0.70\phn \\
\phn2 & HD 189733\dotfill   &  0.357~$\pm$~0.020   &  0.3609~$\pm$~0.0055 & 19.7~$\pm$~1.0\phn       &  50.70~$\pm$~2.55\phn & 51.94~$\pm$~0.87\phn \\
\phn3 & HD 209458\dotfill   &  0.2269~$\pm$~0.0083 &  0.2211~$\pm$~0.0038 & 47.4~$\pm$~1.6\phn       &  21.12~$\pm$~0.72\phn & 21.24~$\pm$~1.00\phn \\
\phn4 & OGLE-TR-10\dotfill  &  \nodata             &  \nodata             &   \nodata                &  \nodata              & \nodata          \\
\phn5 & OGLE-TR-56\dotfill  &  \nodata             &  \nodata             &   \nodata                &  \nodata              & \nodata          \\
\phn6 & OGLE-TR-111\dotfill &  \nodata             &  \nodata             &   \nodata                &  \nodata              & \nodata          \\
\phn7 & OGLE-TR-113\dotfill &  \nodata             &  \nodata             &   \nodata                &  \nodata              & \nodata          \\
\phn8 & OGLE-TR-132\dotfill &  \nodata             &  \nodata             &   \nodata                &  \nodata              & \nodata          \\
\phn9 & TrES-1\dotfill      &  0.0493~$\pm$~0.0024 &  0.0487~$\pm$~0.0006 &  152.3~$\pm$~6.7\phn\phn &   6.57~$\pm$~0.29     & \nodata          \\
10 & TrES-2\dotfill         &  0.0439~$\pm$~0.0028 &  0.0451~$\pm$~0.0007 &  213~$\pm$~11\phn        &   4.70~$\pm$~0.25     & \nodata          \\
11 & TrES-3\dotfill         &  0.0309~$\pm$~0.0018 &  0.0339~$\pm$~0.0004 &  245~$\pm$~13\phn        &   4.08~$\pm$~0.22     & \nodata          \\
12 & TrES-4\dotfill         &  0.0349~$\pm$~0.0025 &  0.0352~$\pm$~0.0005 &  485~$\pm$~31\phn        &   2.06~$\pm$~0.13     & \nodata          \\
13 & WASP-1\dotfill         &  0.0346~$\pm$~0.0026 &  0.0359~$\pm$~0.0005 &  408~$\pm$~27\phn        &   2.45~$\pm$~0.16     & \nodata          \\
14 & WASP-2\dotfill         &  0.0471~$\pm$~0.0096 &  0.0560~$\pm$~0.0013 &  166~$\pm$~30\phn        &   6.03~$\pm$~1.10     & \nodata          \\
15 & XO-1\dotfill           &  0.0506~$\pm$~0.0041 &  0.0531~$\pm$~0.0007 &  172~$\pm$~12\phn        &   5.82~$\pm$~0.42     & \nodata          \\
16 & XO-2\dotfill           &  0.0610~$\pm$~0.0045 &  0.0609~$\pm$~0.0009 &  148~$\pm$~10\phn        &   6.75~$\pm$~0.46     & \nodata          \\
17 & HAT-P-1\dotfill        &  0.0680~$\pm$~0.0071 &  0.0699~$\pm$~0.0011 &  155~$\pm$~15\phn        &   6.44~$\pm$~0.61     & \nodata          \\
18 & HAT-P-2\dotfill        &  0.127~$\pm$~0.018   &  0.1169~$\pm$~0.0016 &  110~$\pm$~13\phn        &   9.07~$\pm$~1.06     &  7.39~$\pm$~0.88 \\
19 & HAT-P-3\dotfill        &  0.0559~$\pm$~0.0059 &  0.0594~$\pm$~0.0011 &  139~$\pm$~13\phn        &   7.21~$\pm$~0.68     & \nodata          \\
20 & HAT-P-4\dotfill        &  0.0474~$\pm$~0.0042 &  0.0457~$\pm$~0.0007 &  314~$\pm$~24\phn        &   3.19~$\pm$~0.24     & \nodata          \\
21 & HAT-P-5\dotfill        &  0.0317~$\pm$~0.0031 &  0.0334~$\pm$~0.0005 &  342~$\pm$~30\phn        &   2.93~$\pm$~0.26     & \nodata          \\
22 & HAT-P-6\dotfill        &  0.0522~$\pm$~0.0046 &  0.0534~$\pm$~0.0008 &  261~$\pm$~20\phn        &   3.83~$\pm$~0.29     & \nodata          \\
23 & GJ 436\dotfill         &  \nodata             &  \nodata             &  \nodata                 &   \nodata             & 97.73~$\pm$~2.27\phn \\ [-2.0ex]
\enddata
\tablecomments{$\phi_{\rm mod}$ and $\phi_{\rm SB}$ are the angular
diameters derived using eq.\,(\ref{eq:modelphi}) and
eq.\,(\ref{eq:kervella}), respectively. The distance and parallax are
computed from $M_V$ (Table~\ref{tab:stellar}) and $V$
(Table~\ref{tab:atmospheric}), ignoring extinction.}
\end{deluxetable*}

\clearpage

\begin{landscape}
\begin{deluxetable*}{clcccccccc}
\tabletypesize{\scriptsize}
\tablewidth{0pc}
\tablecaption{Properties of extrasolar transiting planets.\label{tab:planetary}}
\tablehead{
\colhead {} & 
\colhead{} & 
\colhead{$M_p$} & 
\colhead{$R_p$} &
\colhead{$\log g_p$} & 
\colhead{$\rho_p$} & 
\colhead{$a$}& 
\colhead{} & 
\colhead{$T_{\rm eq}$} &
\colhead{$M_{\rm Z}$} \\
\colhead{\#} & 
\colhead{~~~~~~~~~Name~~~~~~~~~} & 
\colhead{(M$_{\rm Jup}$)} &
\colhead{(R$_{\rm Jup}$)} & 
\colhead{(cgs)} & 
\colhead{(g cm$^{-3}$)} & 
\colhead{(AU)} & 
\colhead{$\Theta$} & 
\colhead{(K)} &
\colhead{(M$_{\earth}$)}
}
\startdata
 1 & HD 149026\dotfill   & $0.359_{-0.021}^{+0.022}$ & $0.654_{-0.045}^{+0.060}$ & $3.360_{-0.088}^{+0.044}$ & $ 1.59_{-0.36}^{+0.38}$    & $0.04313_{-0.00056}^{+0.00065}$ & $0.0384_{-0.0037}^{+0.0028}$    &  $1634_{ -23}^{+90}$ & $68_{-12}^{+10}$   \\ [+1.5ex]
 2 & HD 189733\dotfill   & $1.144_{-0.056}^{+0.057}$ & $1.138_{-0.027}^{+0.027}$ & $3.340_{-0.014}^{+0.014}$ & $ 0.963_{-0.079}^{+0.088}$ & $0.03099_{-0.00063}^{+0.00060}$ & $0.0772_{-0.0027}^{+0.0028}$    &  $1201_{ -12}^{+13}$ & 0--35              \\ [+1.5ex]
 3 & HD 209458\dotfill   & $0.685_{-0.014}^{+0.015}$ & $1.359_{-0.019}^{+0.016}$ & $2.963_{-0.005}^{+0.005}$ & $ 0.338_{-0.014}^{+0.016}$ & $0.04707_{-0.00047}^{+0.00046}$ & $0.04234_{-0.00057}^{+0.00058}$ &  $1449_{ -12}^{+12}$ & 0                  \\ [+1.5ex]
 4 & OGLE-TR-10\dotfill  & $0.62_{-0.14}^{+0.14}$    & $1.25_{-0.12}^{+0.14}$    & $3.00_{-0.13}^{+0.11}$    & $ 0.40_{-0.12}^{+0.18}$    & $0.0434_{-0.0015}^{+0.0013}$    & $0.0386_{-0.0085}^{+0.0089}$    &  $1481_{ -55}^{+71}$ & 0--25              \\ [+1.5ex]
 5 & OGLE-TR-56\dotfill  & $1.39_{-0.17}^{+0.18}$    & $1.363_{-0.090}^{+0.092}$ & $3.264_{-0.069}^{+0.066}$ & $ 0.68_{-0.14}^{+0.18}$    & $0.02383_{-0.00051}^{+0.00046}$ & $0.0393_{-0.0050}^{+0.0052}$    &  $2212_{ -63}^{+61}$ & 0                  \\ [+1.5ex]
 6 & OGLE-TR-111\dotfill & $0.55_{-0.10}^{+0.10}$    & $1.051_{-0.052}^{+0.057}$ & $3.088_{-0.092}^{+0.080}$ & $ 0.59_{-0.13}^{+0.16}$    & $0.04689_{-0.00097}^{+0.0010}$  & $0.057_{-0.011}^{+0.011}$       &  $1025_{ -25}^{+26}$ & $10_{-10}^{+20}$   \\ [+1.5ex]
 7 & OGLE-TR-113\dotfill & $1.26_{-0.16}^{+0.16}$    & $1.093_{-0.019}^{+0.028}$ & $3.419_{-0.065}^{+0.057}$ & $ 1.20_{-0.16}^{+0.17}$    & $0.02289_{-0.00015}^{+0.00016}$ & $0.0677_{-0.0087}^{+0.0086}$    &  $1341_{ -25}^{+30}$ & $25_{-25}^{+10}$   \\ [+1.5ex]
 8 & OGLE-TR-132\dotfill & $1.18_{-0.13}^{+0.14}$    & $1.20_{-0.11}^{+0.15}$    & $3.276_{-0.084}^{+0.10}$  & $ 0.85_{-0.26}^{+0.32}$    & $0.03035_{-0.00053}^{+0.00057}$ & $0.0440_{-0.0058}^{+0.0068}$    &  $2013_{-108}^{+77}$ & $5_{-5}^{+85}$     \\ [+1.5ex]
 9 & TrES-1\dotfill      & $0.752_{-0.046}^{+0.047}$ & $1.067_{-0.021}^{+0.022}$ & $3.220_{-0.026}^{+0.024}$ & $ 0.769_{-0.064}^{+0.069}$ & $0.03925_{-0.00060}^{+0.00056}$ & $0.0634_{-0.0036}^{+0.0036}$    &  $1140_{ -12}^{+13}$ & $20_{-20}^{+20}$   \\ [+1.5ex]
10 & TrES-2\dotfill      & $1.200_{-0.053}^{+0.051}$ & $1.224_{-0.041}^{+0.041}$ & $3.298_{-0.016}^{+0.016}$ & $ 0.813_{-0.083}^{+0.096}$ & $0.03558_{-0.00077}^{+0.00070}$ & $0.0709_{-0.0021}^{+0.0022}$    &  $1498_{ -17}^{+17}$ & 0                  \\ [+1.5ex]
11 & TrES-3\dotfill      & $1.938_{-0.063}^{+0.062}$ & $1.312_{-0.041}^{+0.033}$ & $3.452_{-0.022}^{+0.022}$ & $ 1.065_{-0.085}^{+0.11}$  & $0.02272_{-0.00026}^{+0.00017}$ & $0.0738_{-0.0026}^{+0.0026}$    &  $1623_{ -25}^{+26}$ & 0--30              \\ [+1.5ex]
12 & TrES-4\dotfill      & $0.920_{-0.072}^{+0.073}$ & $1.751_{-0.062}^{+0.064}$ & $2.872_{-0.039}^{+0.037}$ & $ 0.213_{-0.027}^{+0.030}$ & $0.05092_{-0.00069}^{+0.00072}$ & $0.0384_{-0.0030}^{+0.0030}$    &  $1785_{ -29}^{+29}$ & 0                  \\ [+1.5ex]
13 & WASP-1\dotfill      & $0.918_{-0.090}^{+0.091}$ & $1.514_{-0.047}^{+0.052}$ & $3.010_{-0.050}^{+0.044}$ & $ 0.328_{-0.043}^{+0.048}$ & $0.03957_{-0.00048}^{+0.00049}$ & $0.0374_{-0.0037}^{+0.0037}$    &  $1811_{ -27}^{+34}$ & 0                  \\ [+1.5ex]
14 & WASP-2\dotfill      & $0.915_{-0.093}^{+0.090}$ & $1.071_{-0.083}^{+0.080}$ & $3.287_{-0.033}^{+0.038}$ & $ 0.92_{-0.20}^{+0.27}$    & $0.03138_{-0.00154}^{+0.00130}$ & $0.0596_{-0.0041}^{+0.0046}$    &  $1304_{ -54}^{+54}$ & $20_{-15}^{+85}$   \\ [+1.5ex]
15 & XO-1\dotfill        & $0.918_{-0.078}^{+0.081}$ & $1.206_{-0.042}^{+0.047}$ & $3.211_{-0.043}^{+0.037}$ & $ 0.650_{-0.086}^{+0.096}$ & $0.04928_{-0.00099}^{+0.00089}$ & $0.0744_{-0.0061}^{+0.0061}$    &  $1196_{ -19}^{+23}$ & 0--20              \\ [+1.5ex]
16 & XO-2\dotfill        & $0.566_{-0.055}^{+0.055}$ & $0.983_{-0.028}^{+0.029}$ & $3.168_{-0.047}^{+0.043}$ & $ 0.741_{-0.091}^{+0.10}$  & $0.03684_{-0.00043}^{+0.00040}$ & $0.0438_{-0.0042}^{+0.0042}$    &  $1319_{ -23}^{+24}$ & $30_{-15}^{+15}$   \\ [+1.5ex]
17 & HAT-P-1\dotfill     & $0.532_{-0.030}^{+0.030}$ & $1.242_{-0.053}^{+0.053}$ & $2.931_{-0.025}^{+0.025}$ & $ 0.345_{-0.044}^{+0.053}$ & $0.0553_{-0.0013}^{+0.0012}$    & $0.0418_{-0.0019}^{+0.0020}$    &  $1306_{ -30}^{+30}$ & 0                  \\ [+1.5ex]
18 & HAT-P-2\dotfill     & $8.72_{-0.36}^{+0.39}$    & $1.003_{-0.066}^{+0.084}$ & $4.370_{-0.078}^{+0.019}$ & $10.7_{-2.3}^{+2.5}$       & $0.0679_{-0.0014}^{+0.0015}$    & $0.941_{-0.075}^{+0.035}$       &  $1398_{ -33}^{+61}$ & \nodata            \\ [+1.5ex]
19 & HAT-P-3\dotfill     & $0.596_{-0.026}^{+0.024}$ & $0.899_{-0.049}^{+0.043}$ & $3.310_{-0.072}^{+0.059}$ & $ 1.02_{-0.14}^{+0.19}$    & $0.03882_{-0.00077}^{+0.00060}$ & $0.0585_{-0.0048}^{+0.0044}$    &  $1127_{ -39}^{+49}$ & $70_{-35}^{+25}$   \\ [+1.5ex]
20 & HAT-P-4\dotfill     & $0.671_{-0.044}^{+0.033}$ & $1.274_{-0.060}^{+0.049}$ & $3.020_{-0.023}^{+0.016}$ & $ 0.403_{-0.049}^{+0.065}$ & $0.04438_{-0.0015}^{+0.00081}$  & $0.0378_{-0.0014}^{+0.0016}$    &  $1686_{ -26}^{+30}$ & 0                  \\ [+1.5ex]
21 & HAT-P-5\dotfill     & $1.06_{-0.11}^{+0.11}$    & $1.254_{-0.056}^{+0.051}$ & $3.219_{-0.051}^{+0.047}$ & $ 0.66_{-0.10}^{+0.12}$    & $0.04071_{-0.00097}^{+0.00049}$ & $0.0591_{-0.0062}^{+0.0064}$    &  $1539_{ -32}^{+33}$ & 0                  \\ [+1.5ex]
22 & HAT-P-6\dotfill     & $1.059_{-0.052}^{+0.053}$ & $1.330_{-0.058}^{+0.064}$ & $3.171_{-0.030}^{+0.029}$ & $ 0.559_{-0.076}^{+0.086}$ & $0.05237_{-0.00090}^{+0.00085}$ & $0.0646_{-0.0031}^{+0.0032}$    &  $1675_{ -31}^{+32}$ & 0                  \\ [+1.5ex]
23 & GJ 436\dotfill      & $0.0729_{-0.0025}^{+0.0025}$ & $0.3767_{-0.0092}^{+0.0082}$ & $3.107_{-0.041}^{+0.039}$ & $ 1.69_{-0.12}^{+0.14}$ & $0.02872_{-0.00026}^{+0.00029}$    & $0.0246_{-0.0013}^{+0.0013}$ &  $649_{-60}^{+60}$\phn          & \nodata            \\ [-1.0ex]
\enddata
\tablecomments{$\Theta$ is the Safronov number, $T_{\rm eq}$ the
zero-albedo equilibrium temperature ignoring the energy
redistribution factor, and $M_{\rm Z}$ is the planet heavy element
content.}
\end{deluxetable*}
\clearpage
\end{landscape}

\end{document}